\documentclass[apjl,iop]{emulateapj}

\usepackage{amssymb,natbib,graphicx,subfigure,color,apjfonts}

\def\Hb{{\rm H\beta}}

\def\SUBFIND{{\small SUBFIND}}

\def\SFR{{\rm SFR}}

\newcommand{\ifm}[1]{\relax\ifmmode#1\else$\mathsurround=0pt #1$\fi}
\newcommand{\be}{\begin{equation}}
\newcommand{\ee}{\end{equation}}
\newcommand{\bea}{\begin{eqnarray}}
\newcommand{\eea}{\end{eqnarray}}

\newcommand{\Fig}[1]{Figure (\ref{f:#1})}
\newcommand{\Figs}[2]{Figures (\ref{f:#1}) and (\ref{f:#2})}

\newcommand{\sggg}[1]{\textcolor{green}{[]}}

\def\dex{{\rm\thinspace dex}}
\def\cm{{\rm\thinspace cm}}

\def\pc{{\rm\thinspace pc}}
\def\kpc{{\rm\thinspace kpc}}
\def\Mpc{{\rm\thinspace Mpc}}

\def\Msun{\hbox{$\rm\thinspace M_{\odot}$}}
\def\Zsun{\hbox{$\rm\thinspace Z_{\odot}$}}
       
\def\yr{{\rm\thinspace yr}}

\def\Msunyr{\Msun\yr^{-1}}
\def\Msunpc2{{\Msun\pc}^{-2}}
\def\Msunyrkpc2{{\Msun\yr^{-1}\kpc}^{-2}}

\def\magarcsec2{{\rm\thinspace mag\thinspace arcsec}^{-2}}

\slugcomment{The Astrophysical Journal, accepted}

\shorttitle{Metallicity and Environment in Illustris}
\shortauthors{Genel, S.}

\begin{document}

\title{How Environment Affects Galaxy Metallicity through Stripping and Formation History: Lessons from the Illustris Simulation}

\author{Shy Genel\altaffilmark{1,2}}

\altaffiltext{1}{Department of Astronomy, Columbia University, 550 West 120th Street, New York, NY 10027, USA}
\altaffiltext{2}{Hubble Fellow}
\email{shygenelastro@gmail.com}

\begin{abstract}
Recent studies have found higher galaxy metallicities in richer environments. It is not yet clear, however, whether metallicity-environment dependencies are merely an indirect consequence of environmentally dependent formation histories, or of environment related processes directly affecting metallicity. Here, we present a first detailed study of metallicity-environment correlations in a cosmological hydrodynamical simulation, in particular the Illustris simulation. Illustris galaxies display similar relations to those observed. Utilizing our knowledge of simulated formation histories, and leveraging the large simulation volume, we construct galaxy samples of satellites and centrals that are matched in formation histories. This allows us to find that $\sim1/3$ of the metallicity-environment correlation is due to different formation histories in different environments. This is a combined effect of satellites (in particular, in denser environments) having on average lower $z=0$ star formation rates (SFRs), and of their older stellar ages, even at a given $z=0$ SFR. Most of the difference, $\sim2/3$, however, is caused by the higher concentration of star-forming disks of satellite galaxies, as this biases their SFR-weighted metallicities toward their inner, more metal-rich parts. With a newly defined quantity, the `radially averaged' metallicity, which captures the metallicity profile but is independent of the SFR profile, the metallicities of satellites and centrals become environmentally independent once they are matched in formation history. We find that circumgalactic metallicity (defined as rapidly inflowing gas around the virial radius), while sensitive to environment, has no measurable effect on the metallicity of the star-forming gas inside the galaxies.
\end{abstract}


\section{Introduction}
\label{s:intro}
The metal content of galaxies encodes important information about their formation histories and the physical processes governing them. It is observationally well-established that galaxies with higher stellar masses tend to have higher metallicities \citep{LequeuxJ_79a,TremontiC_04a,GallazziA_05}, and also that the gas-phase metallicity of galaxies with higher star formation rates (SFRs) or gas fractions tends to be {\it lower} at a given stellar mass \citep{EllisonS_08a,ZhangW_09a,MannucciF_10a,BothwellM_13a}. These correlations are generally understood as resulting from the interplay between metal production by stars and dilution due to the accretion of metal-poor intergalactic gas \citep{TinsleyB_80a,FinlatorK_08a,DaveR_12a,ZahidJ_14a}. Galactic outflows are also believed to play an important role by modifying both the galactic gas and metal contents.

However, large systematic uncertainties, both observationally and theoretically, severely limit the constraining power of these observed correlations on theoretical models. The normalization of metallicity measurements depends on the calibration scheme, with differences between different schemes that are comparable to the variation among different galaxy populations \citep{KewleyL_08a,NewmanS_14a}. Moreover, even relative metallicities suffer from systematics, such that the quantitative nature of correlations such as those described above is uncertain even if the qualitative trends are mostly robust \citep{KewleyL_08a,AndrewsB_13a,Perez-MonteroE_13a,delosReyesM_15a}. Issues also exist on the theoretical side, pertaining to large uncertainties regarding stellar populations and their metal yields \citep{CoteB_15a,VincenzoF_16a}. These limitations motivate the expansion of the parameter space of the galaxy properties with which galaxy metallicities should be compared, in the hope that additional qualitative trends - in the absence of robust quantitative relations - will reveal themselves as containing independent constraining power.

An important extension of the parameter space beyond the stellar mass and SFR is the galaxy environment. Several studies in the past few years have found that galaxies in `richer' environments tend to have higher gas-phase metallicities \citep{MouhcineM_07a,CooperM_08a}. The exact magnitude of the effect, and the determining environmental factor, are still a matter of debate. Some studies found that satellite galaxies have higher metallicities than central galaxies of the same stellar mass, and that the effect is stronger for larger host halo masses and smaller halo-centric distances \citep{PasqualiA_12a,PetropoulouV_12a}. Some studies deduced that the underlying correlation is to environmental overdensity, with no significance as to whether the galaxy is a central or satellite galaxy \citep{EllisonS_09a}. \citet{PengY_14a} find that metallicity correlates with environmental overdensity primarily for satellite galaxies, but much less so for central galaxies. While the tendency of galaxies in richer environments to have higher metallicities is generally agreed upon at $z\approx0$, the measured trends at higher redshifts span both positive and negative correlations \citep{MagriniL_12a,KulasK_13a,LaraLopezM_13d,DarvishB_15a,KacprzakG_15a,ShimakawaR_15a,ValentinoF_15a}, and the verdict is still out.

The question remains open concerning the physical processes that drive the observed correlations. Can the environmental dependencies be fully explained by the different star formation histories of satellite and central galaxies, which are driven by the environmental processes, and the `standard' relations between star formation and metallicity \citep{HughesT_13a}? Or do the more direct influences of environment on galactic metallicity, e.g.~through the metallicity of enriched intergalactic gas accreted onto galaxies, play a more significant role \citep{RobertsonP_12a,PengY_14a}? Through its sensitivity to the myriad of processes that control the gas content of galaxies, metallicity has the potential to teach us about the environmental sensitivity of `standard' gas processes, such as consumption and recycling, as well as environment-specific processes, such as gas stripping and strangulation. Several environment-specific scenarios have been proposed in the literature to explain the observed trends (see, e.g., the discussion in \citealp{PetropoulouV_12a}). These scenarios include the possibilities that satellites experience {\bf (i)} more enriched accretion from their environments relative to centrals, {\bf (ii)} suppression of accretion from a low-metallicity reservoir due to stripping and strangulation, or {\bf (iii)} suppression of winds that allow metals to escape, i.e.~increased wind recycling. All of these scenarios are expected to be more effective in more massive host halos and denser environments (e.g.~closer to the host halo center), in agreement with observations.

Detailed analysis of simulations is an invaluable tool for distinguishing between these scenarios, as well as for finding new ones, as is done in this work. Less than a handful of cosmological hydrodynamical simulations have thus far been used to investigate the metallicity-environment relations. At least qualitatively, they agree with the observational results \citep{DaveR_11b,DeRossiM_15a}. However, the emerging correlations have not yet been explained. Here, we exploit the power of the Illustris simulation \citep{GenelS_14a,SijackiD_14a,VogelsbergerM_14a,VogelsbergerM_14b}, both in terms of galaxy population size and of physical fidelity, to explore in detail, for the first time, the metallicity-environment relations in a population of simulated $z=0$ galaxies. In Section \ref{s:methods} we describe the methods used in this paper. In Section \ref{s:Z_SFR} we present a short account of the anti-correlation between metallicity and SFR in Illustris that serves as a reference for the following sections. In Section \ref{s:obscomp} we show that satellite galaxies in Illustris tend to have higher metallicities than central galaxies and, in Section \ref{s:SFRweighted}, we show that this can be partially explained by their different formation histories. In Section \ref{s:resolvedZ}, however, we show that most of the difference is due to different SFR profiles between the two populations, which are a result of environmental effects. In Section \ref{s:envir} we quantify the metallicity in relation to environmental overdensity, and show that the same two factors discussed in Section \ref{s:cent_sat} can also explain these trends. In Section \ref{s:summary}, we summarize our results.

\section{Methods}
\label{s:methods}
The Illustris simulation is a hydrodynamical cosmological simulation of structure formation that follows a comoving volume of $(106.5\Mpc)^3$ from $z=127$ to $z=0$ with a baryonic mass resolution of $\approx1.6\times10^6\Msun$. In this study, we focus on galaxies in a particular narrow stellar mass bin of $M_*=(2-3)\times10^{10}\Msun$, in the interest of simplicity and because the mass-metallicity relation itself is not our focus; however, we note that our conclusions qualitatively apply to the full range of $M_*\sim10^9-10^{11}\Msun$. The galaxies in our selected narrow mass bin are resolved in the Illustris simulation with $\approx3\times10^4$ collisionless particles for the stellar component, and with typically one order of magnitude fewer hydrodynamical cells representing the disk of star-forming gas. In star-forming disks at $z=0$, the gravitational softening and hydrodynamical resolution (typical cell diameter) are comparable at $\approx700\pc$. In Sections \ref{s:cent_sat} and \ref{s:envir} we distinguish between central and satellite galaxies as defined by \SUBFIND{ }\citep{SpringelV_01}. We utilize information on the formation histories of individual galaxies by following the main-progenitor branches of the galaxy merger trees of \citet{Rodriguez-GomezV_14a}. We quantify the environment of each galaxy by measuring the distance to its fifth-nearest bright galaxy and converting those distances to environmental overdensities, $\delta$ \citep{VogelsbergerM_14b}.

The physical processes implemented in the Illustris simulation, laid out in detail in \citet{VogelsbergerM_13a}, include gravity, hydrodynamics, and radiative cooling, which form the backbone of galaxy formation and, in particular, have great potential to drive environmentally dependent effects. These components of the physical model are relatively robust and numerically accurate \citep{SpringelV_10a,VogelsbergerM_12a,VogelsbergerM_13a}. Also, crucially for generating semi-realistic galaxy populations, there are sub-grid models for star formation, black hole growth, and associated feedback processes.

It is worth reminding the reader of the most relevant aspects of the physical models. {\bf (i)} In order to perform direct comparisons to observations of galactic gas-phase metallicities, which are obtained by probing emission-line ratios, we study the metallicity of star-forming gas. All of the gas in the simulation with a hydrogen number density of $n>0.13\cm^{-3}$ is defined as star forming, and is assigned an SFR according to a \citet{SchmidtM_59a} law \citep{SpringelV_03a,VogelsbergerM_13a}. We consider and contrast both the SFR-weighted metallicities and the radially averaged metallicities, which are described in the following sections. {\bf (ii)} Metals are produced by evolving stellar populations, represented in Illustris by colliosionless stellar particles. These particles release mass (hydrogen, helium, and seven additional species of heavier elements) into their surrounding gas in a continuous fashion, using tabulated rates determined by the age of the stellar particle and its initial metallicity. In this paper, we do not distinguish between the various heavy elements that the simulation follows, but define the units of metallicity as the mass fraction in all of the elements heavier than helium, in logarithmic units with a basis of $10$. In these units, the solar metallicity is $-1.85$ \citep{AsplundM_09a}. {\bf (iii)} Since we are concerned with the environmental metallicity around galaxies, it is worth mentioning that metals are ejected outside of galaxies in our model primarily by two feedback processes. First, galactic winds, whose mass-loading is a decreasing function of galaxy mass, and whose ejection velocities scale with, and are somewhat below, the escape velocity from the halo. These winds expel metals out of the galaxies where they were produced but not significantly beyond their halos, allowing them instead to participate in a `halo fountain' that is driven by the hydrodynamics in the halo gas and the cooling rate \citep{NelsonD_15a,SureshJ_14a}. Second, feedback from black holes dominates the metal ejection out of halos into the cosmic web \citep{SureshJ_14a}.

\section{Background result: metallicity and SFR}
\label{s:Z_SFR}

\begin{figure*}
\centering
\includegraphics[width=1.0\textwidth]{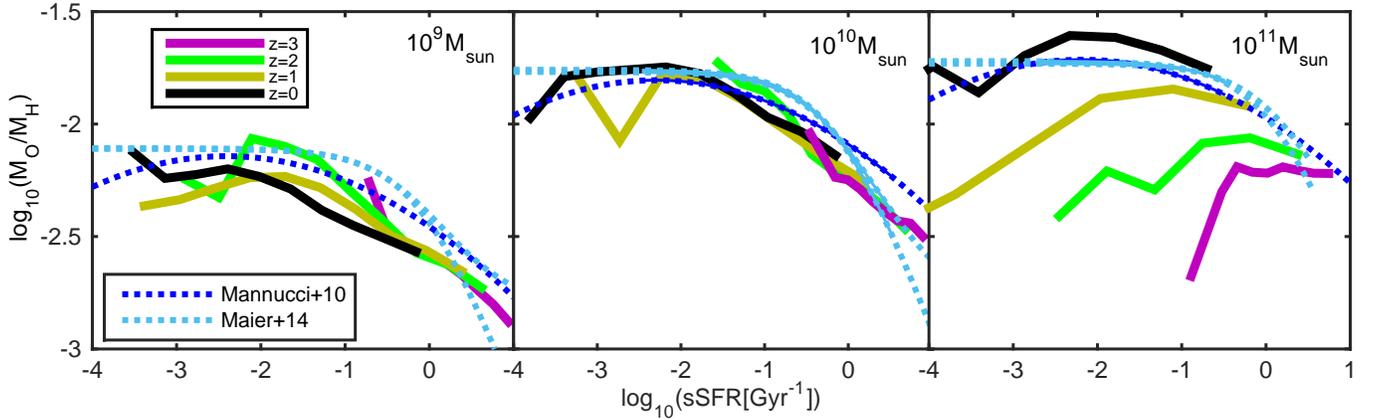}
\caption{Three cuts, for different stellar masses (different panels), of the fundamental metallicity relation (FMR). Observational inferences of the relation \citep{MannucciF_10a,MaierC_14a} are shown in blue (dark and light, respectively), and include both the regimes where they were actually constrained observationally (thin solid), and extrapolations of the respective fitting formulae (dotted). Results from Illustris are shown for different redshifts $z=0,1,2,3$ (thick solid). To make direct comparisons to observations, this is the only plot in this paper where a particular element (oxygen) is used rather than the total metallicity.}
\vspace{0.3cm}
\label{f:Z_SFR}
\end{figure*}

Numerous recent observational studies have found an anti-correlation between metallicity and SFR at various redshifts (e.g.~\citealp{AndrewsB_13a,BelliS_13a,delosReyesM_15a}). Most commonly, these relations are described in terms of the fundamental metallicity relation (FMR), i.e., the relation between gas metallicity, stellar mass, and specific star formation rate \citep{MannucciF_10a}, which is presented in \Fig{Z_SFR}. There is considerable debate in the literature as to whether or not this relation evolves with redshift (e.g.~\citealp{Perez-MonteroE_13a,SteidelC_14a,WuytsE_14a,SandersR_15a}). For example, \citet{MaierC_14a} offer an alternative formulation of the FMR that is based on their measurements, and claim that in this form \citep{LillyS_13a}, rather than in the original \citet{MannucciF_10a} form, it is indeed epoch-independent.

We find that Illustris reproduces these observations quite well given the observational uncertainties, in particular, in terms of the trend with specific SFR (sSFR) and the very weak FMR evolution with redshift. An exception to this last point is a strong evolution for massive galaxies of $M_*\approx10^{11}\Msun$. While at high sSFR the relations at all redshifts converge toward the extrapolation of the observed relation, including at this high mass bin, at low sSFR they diverge, showing lower metallicities at higher redshifts. This is the regime where AGN feedback controls the mass-metallicity-sSFR relation, breaking the FMR, as will be discussed in future work. The emergence of an FMR-like relation is a rather generic result in cosmological hydrodynamical simulations \citep{DaveR_11b,DeRossiM_15a,LagosC_16a}. It is not the focus of our current work, and is shown here only as a reference, since the general anti-correlation between metallicity and SFR will come up several times in the following sections.

\section{Main results: centrals versus satellites}
\label{s:cent_sat}
\subsection{Comparison to observations}
\label{s:obscomp}
\begin{figure*}
\centering
\subfigure[]{
          \label{f:MZrelation_fullextent}
          \includegraphics[width=0.49\textwidth]{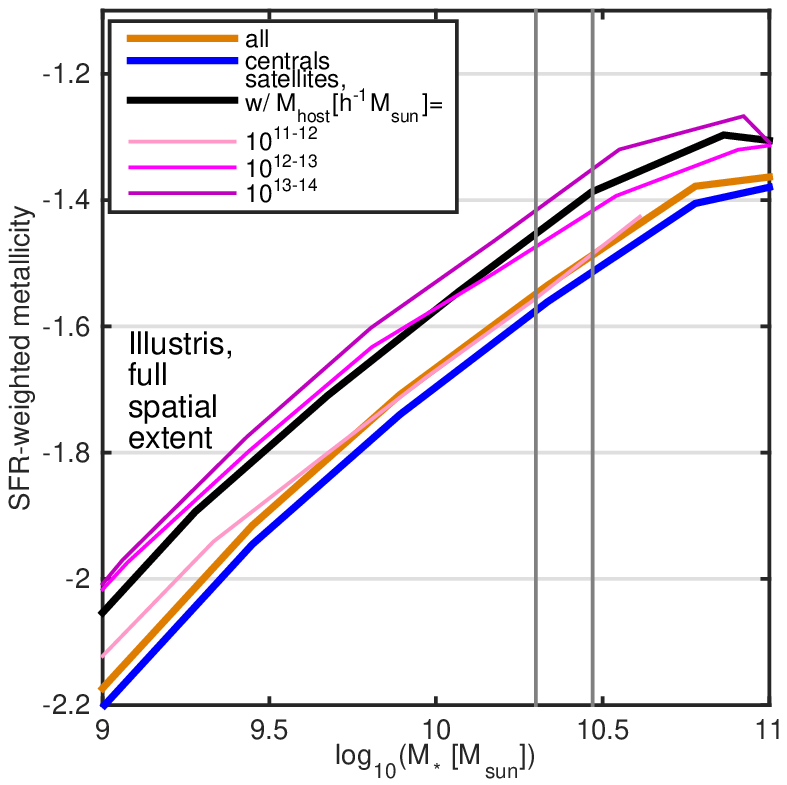}}
\subfigure[]{
          \label{f:MZrelation_5kpc}
          \includegraphics[width=0.49\textwidth]{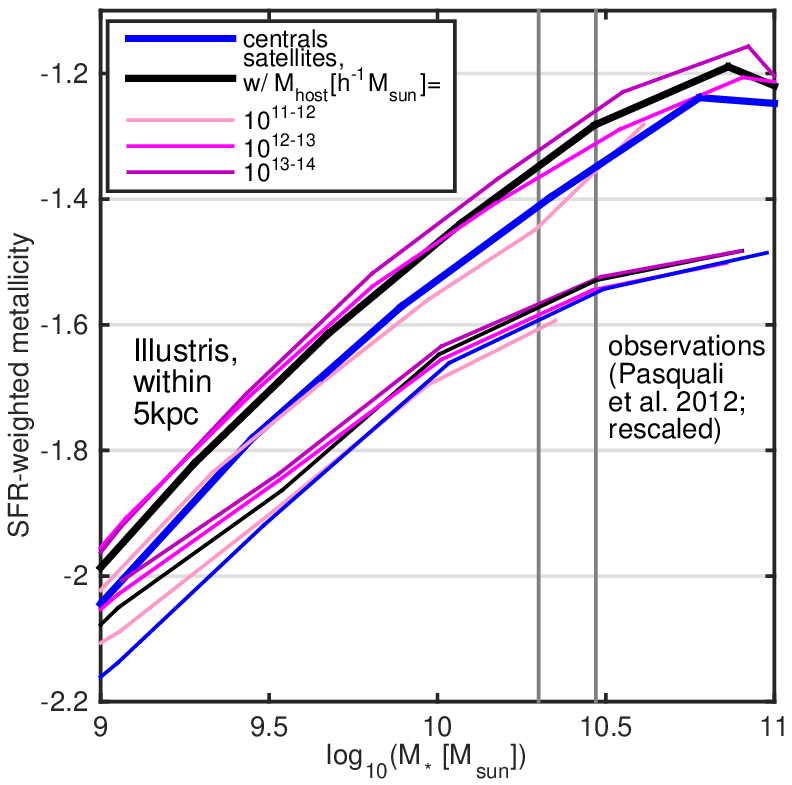}}
\caption{Relation between median SFR-weighted gas-phase metallicity and stellar mass. Galaxies are distinguished by their group membership: all galaxies (orange), central galaxies (blue), and satellite galaxies (black). The satellites are further separated by the mass of the dark matter halo inside which they are hosted (thin magenta), as indicated in the legend. At a given stellar mass, satellite galaxies tend to have higher metallicities than central galaxies. The effect is stronger for satellites hosted in more massive hosts. Vertical lines indicate the mass range that is the focus of this paper. Left: metallicities within the full spatial extent of the galaxies. Right: metallicities limited to the inner $5\kpc$ of the galaxies (Illustris; upper, thick curves) or to the SDSS fiber (observational results from \citet{PasqualiA_12a}, re-normalized on the y-axis; lower, thin curves).}
\vspace{0.3cm}
\label{f:MZrelation}
\end{figure*}

\Fig{MZrelation} shows the $z=0$ mass-metallicity relation of galaxies in the Illustris simulation. The simulated metallicities are those of the star-forming gas either in the full extent of the galaxies (left panel) or within their inner $5\kpc$ (right panel; upper set of curves), and they are weighted by SFR, analogous to observational measurements. \Fig{MZrelation_5kpc} also presents the observational result of \citet{PasqualiA_12a}, which is based on SDSS and therefore is limited to the inner parts of the galaxies as determined by the SDSS fiber aperture (lower set of curves).

We begin by discussing the qualitative trends seen in \Fig{MZrelation}. More massive Illustris galaxies have higher metallicities. Observations show the same qualitative trend, but quantitatively the simulated relation appears too steep, as pointed out and discussed in detail in \citet{TorreyP_14a}. Here, we also separate the full galaxy population (orange) into central galaxies (blue) and satellites (black). Across more than two orders of magnitude in stellar mass, satellite galaxies have higher metallicities than central galaxies of the same mass. This is in qualitative agreement with observational findings \citep{PasqualiA_12a,PetropoulouV_12a}. Furthermore, we separate the satellite galaxies according to the dark matter halo mass in which they are hosted (magenta), showing that at a given stellar mass, a satellite galaxy tends to have a higher metallicity the more massive its host halo is. This trend too qualitatively agrees with observations. This rough agreement with observations serves as the starting point for our study of how this difference emerges in Illustris.

Quantitatively, the metallicity difference between centrals and satellites, and its sensitivity to the host halo mass, depends on the details of the measurement. Comparing \Figs{MZrelation_fullextent}{MZrelation_5kpc}, it can be seen that the differences are larger for the full spatial extent of the galaxies than in the case where only the metallicities of the inner regions are considered. For the full galaxy extents, the difference decreases from $\approx0.15\dex$ at $M_*=10^9\Msun$ to $\approx0.1\dex$ at $M_*=10^{11}\Msun$. The difference inside $5\kpc$ is in turn $\approx0.06\dex$ at $M_*=10^9\Msun$ and $\approx0.03\dex$ at $M_*=10^{11}\Msun$. At the mass scale on which we focus in this paper, which is indicated by the vertical lines in \Fig{MZrelation}, the differences are $\approx0.13\dex$ and $\approx0.06\dex$ for the full extent and within $5\kpc$, respectively. The effect of the aperture on the dependence of the metallicity on environment is discussed in great detail in Sections \ref{s:resolvedZ} and \ref{s:envir}, a discussion that constitutes a fundamental part of this work. Here, it is merely noted, and the main goal is to perform a comparison with observations.

At face value, \Fig{MZrelation_5kpc} indicates differences that are weaker in the SDSS-based \citet{PasqualiA_12a} observations by about a factor of two compared to Illustris. However, this quantitative comparison should be taken with caution, as a fully robust comparison would require a significantly more complex and complete analysis. We have chosen a value of $5\kpc$ in order to approximately match the aperture effect of the SDSS fiber. The redshifts of the galaxies used by \citet{PasqualiA_12a} are in the range $0.01<z<0.2$, providing a range of physical apertures of $\approx0.6-14\kpc$ corresponding to the $3''$ fiber aperture. The metallicities for these galaxies have been calculated by \citet{TremontiC_04a}, who note that the fraction of galaxy light falling in the fiber aperture is typically about $1/3$. For our selected mass bin of $M_*=(2-3)\times10^{10}\Msun$, $1/3$ of the total star formation resides within $\approx5\kpc$, which indeed roughly matches the median redshift of the observational sample of $z\sim0.1$ \citep{TremontiC_04a}. However, this is not an exact match, of course, but only an average one, and therefore may include some systematic bias. In addition, the observational sample constitutes of only $\sim25\%$ of the parent spectroscopic SDSS sample, mostly due to cuts in the $\Hb$ signal to noise and the removal of AGN contamination \citep{TremontiC_04a,PasqualiA_12a}, both of which preferentially cut out galaxies with low SFRs (for the relevance of the AGN cut, see, e.g.~\citealp{LeslieS_15a}). In contrast, we include all of the Illustris galaxies with non-zero SFRs. These different selections are likely to introduce additional biases that operate in the direction of making the inferred observed effect indeed smaller than the one derived from Illustris, but the exact quantification of these biases is beyond the scope of this paper. While it is important to keep in mind that at face value the environmental effect is stronger in Illustris than is observed, hereafter we continue to perform an analysis of the effect in Illustris, in the hope that it has some bearing on galaxies in the real Universe. In fact, we provide specific observational predictions which, in the near future, will allow one to rule out or corroborate the theoretical scenario we describe.

\subsection{Galaxy-wide SFR-weighted metallicities}
\label{s:SFRweighted}
\begin{figure*}
\centering
\includegraphics[width=1.0\textwidth]{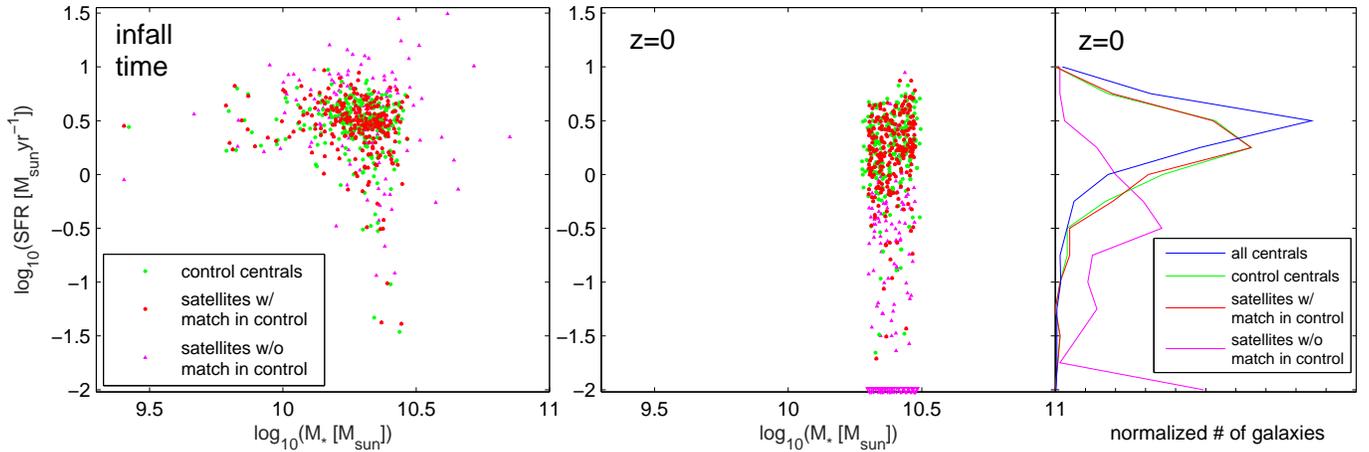}
\caption{Selection criteria for galaxy samples, shown by their joint distributions of SFR and stellar mass at infall time (left panel) and at $z=0$ (middle panel), as well as the $z=0$ SFR distributions collapsed along the mass axis (right panel). The common selection criterion for all galaxies is the narrow mass bin $M_*=(2-3)\times10^{10}\Msun$ at $z=0$. The `matched satellites' sample (red) is the subset of satellites that have a closely matching central galaxy in terms of stellar mass and SFR, both at (their own individual) infall time and at $z=0$. Those matched centrals in turn comprise the `control centrals' sample (green), whose SFR distribution at $z=0$ is shifted only slightly toward lower values compared to the full central sample (blue). The `non-matched satellites' sample (magenta) is comprised of the satellites for which no such close match can be found. Data at $\SFR=0.01\Msunyr$ represent galaxies that in fact have $\SFR=0$.}
\vspace{0.3cm}
\label{f:samples}
\end{figure*}

A principal tool we use throughout the analysis is the comparison of several galaxy samples. As mentioned in Section \ref{s:methods}, we focus on the mass range $M_*=(2-3)\times10^{10}\Msun$, within which we first select the `satellite' and `central' samples, with no additional criteria. At $z=0$, there are $775$ central galaxies inside the mass bin of interest, and $382$ satellite galaxies; these numbers are large enough to allow us to make statistically robust comparisons. The selection criteria for the samples defined in the following are shown in \Fig{samples}. In the next step, for each galaxy in the `satellite' sample, a $z=0$ galaxy from the `central' sample that matches its stellar mass (to within $0.05\dex$) and SFR (to within $0.1\dex$) both at $z=0$ and at its infall time, i.e.~the time it became a satellite, is selected into a `control' sample (green). The $252$ satellite galaxies for which a match exists in the `control' sample comprise the `matched satellites' sample (red), and the remaining $130$, for which a close match could not be found, make up the `non-matched satellites' sample (magenta). \Fig{samples} shows that the `non-matched' satellites tend to have higher SFRs and sometimes high masses at their infall time, but low SFRs at $z=0$. These `early' star formation histories are the reason the `non-matched' satellites could not be matched to the formation history of any central galaxy in the simulation. In contrast, the `matched satellites' and their `control centrals' are very closely matched in their formation histories.

It is worth pointing out that the control sample is not comprised of special central galaxies. This can be seen in several ways. First, \Fig{samples} shows that the distribution of its $z=0$ SFRs is only mildly shifted with respect to the full central sample. Second, $687$ central galaxies of the total $775$ ($88.6\%$) may in principle be selected for the control sample. Namely, for $687$ central galaxies there exists a satellite galaxy that has a similar stellar mass (within $0.05\dex$) and a similar SFR (within $0.1\dex$) both at $z=0$ and at the satellite's infall time. Third, among the $252$ galaxies in the control sample, there are $\approx200$ unique centrals (this number is not well defined, since there is some freedom in selecting the control galaxy for any particular satellite, among all those centrals that are close enough in formation history). Namely, it is not the case that a small number of central galaxies serve multiply as matches to a much larger number of satellites. The key to the control sample is that it is a sample of `normal' centrals (as almost $9$ of any $10$ central galaxies {\it may} serve as a control galaxy) that is, however, biased in terms of formation histories with respect to the full central population, such that it matches those of a large fraction of the satellite galaxies (the ones in the `matched satellites' sample).

\begin{figure}
\centering
\includegraphics[width=0.475\textwidth]{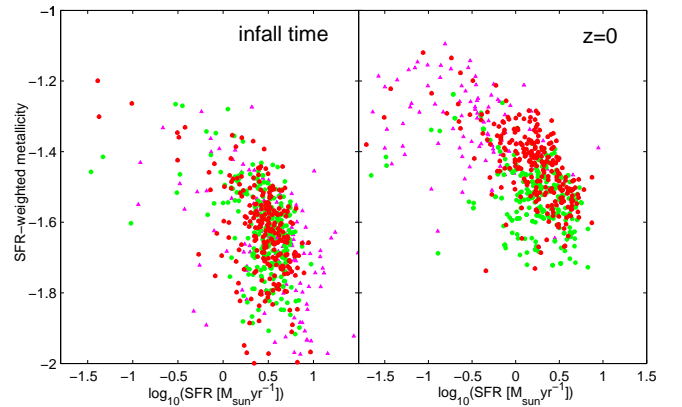}
\caption{Relation between metallicity and SFR for galaxy samples. The `matched satellites' (red) and `control centrals' (green) samples are matched in SFR (and stellar mass, see \Fig{samples}) at both epochs, but while they have statistically indistinguishable metallicities at the infall time, the satellites have significantly higher metallicities ($\approx0.1\dex$) at $z=0$. The `non-matched satellites' (magenta) generally possess lower values of $z=0$ SFR compared to galaxies in the other samples, and correspondingly higher metallicities.}
\vspace{0.3cm}
\label{f:samples_Z}
\end{figure}

Using these samples, we can answer a basic question. When do the high metallicities of $z=0$ satellite galaxies develop? Is it before or only after they become satellites? The answer can be inferred from \Fig{samples_Z}, which displays the relation between metallicity and SFR for the various samples. The relation at infall time (left) appears to be indistinguishable between the two samples that are matched by their formation history. The `non-matched satellites' sample, in turn, has a distinct SFR distribution at the infall time, but also appears to lie on a similar locus in the metallicity-SFR plane. However, the picture is very different at $z=0$ (right) where the satellite galaxies are shown to have markedly higher metallicities, even at a given SFR. Hence, we learn that satellite galaxies only obtain their higher metallicities with respect to centrals after their time of infall.

\begin{figure}
\centering
\subfigure[]{
          \label{f:Mhistory_tinfall}
          \includegraphics[width=0.23\textwidth]{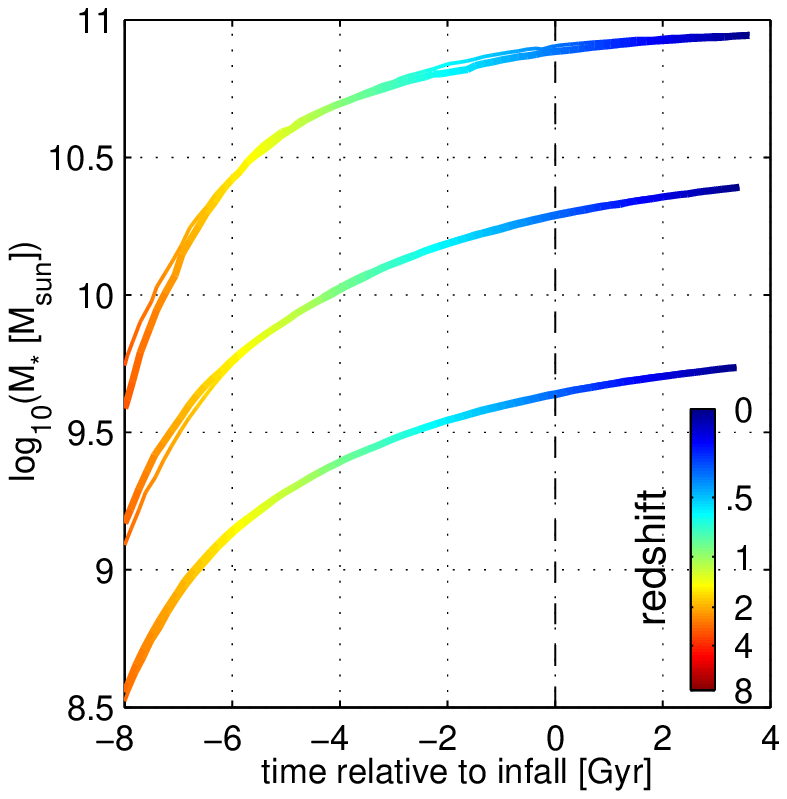}}
\subfigure[]{
          \label{f:Zhistory_tinfall}
          \includegraphics[width=0.23\textwidth]{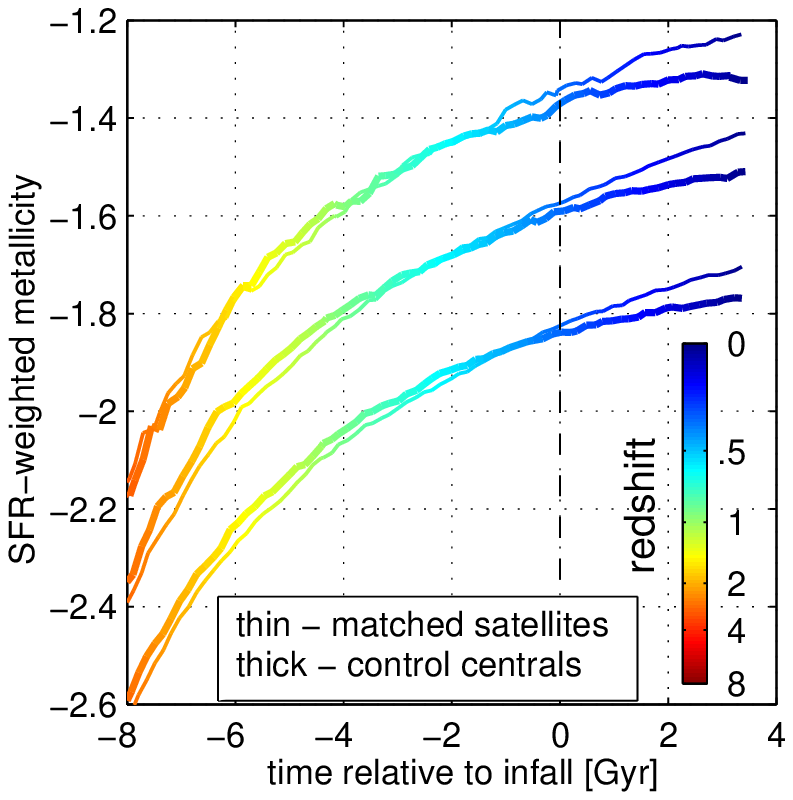}}
\caption{Mean time evolution histories of stellar mass (left) and metallicity (right) for $z=0$ galaxy samples selected in three narrow mass bins. For each mass bin, the thin curves show the `matched satellites' sample, and the thick curves the `control' sample. The x-axis represents time relative to the infall time. Namely, each galaxy is shifted on the x-axis such that the time at which it becomes a satellite is `time zero' (for the `control' galaxies, it is the time the satellite match becomes a satellite). The metallicity growth after infall is much stronger than the mass growth, and can account for most of the metallicity difference between satellites and centrals seen in \Fig{MZrelation}.}
\vspace{0.3cm}
\label{f:history_tinfall}
\end{figure}

\Fig{history_tinfall} provides more details by showing the full mean time evolution\footnote{For a discussion of the validity and pitfalls of the mean formation histories, see \citet{TorreyP_14c}.} of both the mass (left) and metallicity (right), where the time axis is shifted individually for each galaxy such that its infall time is at `time of zero' (marked by a vertical dashed curve). The color indicates the mean redshift along the evolution track, demonstrating that the infall occurs typically at $z\approx0.5$. Two additional mass bins are shown aside from the one that is the focus of this paper (the middle one), demonstrating the same qualitative behavior over a wide mass range. Thin curves show the histories of `matched satellites' and thick curves the histories of `control centrals'. Naturally, the `control centrals' do not have an infall time of their own, as by construction they are still central galaxies at $z=0$. For those, the time axis is the time relative to the infall time of their own (individual) matched satellite galaxy. The mass histories of these samples are almost indistinguishable (for the lowest mass bin, {\it truly} indistinguishable, given the thickness of the curves in \Fig{history_tinfall}) already many gigayears before the infall time, even though they were only matched at $z=0$ and at the infall time itself. This similarity for most of cosmic time renders the comparison between these two samples powerful, as in terms of formation history they only differ by their group membership at late times, not by their mass growth histories. In contrast, the metallicity histories, while being very similar before infall, start diverging right around infall. The difference that develops between infall time and $z=0$ accounts for most of the difference between satellites and centrals seen in \Fig{MZrelation}.

\begin{figure*}
\centering
\subfigure[]{
          \label{f:MZhistory_zcolors}
          \includegraphics[width=0.49\textwidth]{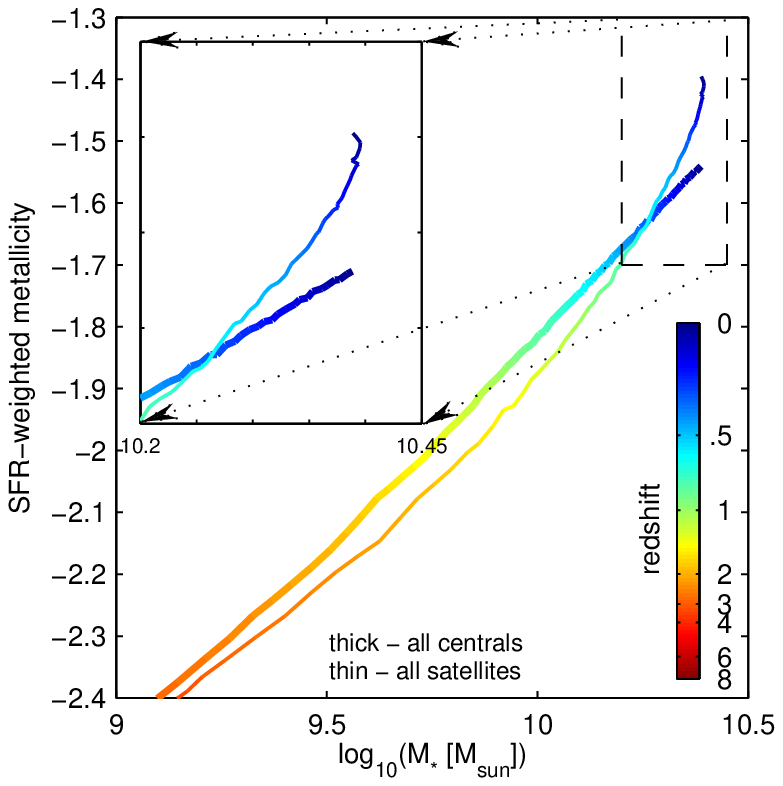}}
\subfigure[]{
          \label{f:MZhistory_samples}
          \includegraphics[width=0.49\textwidth]{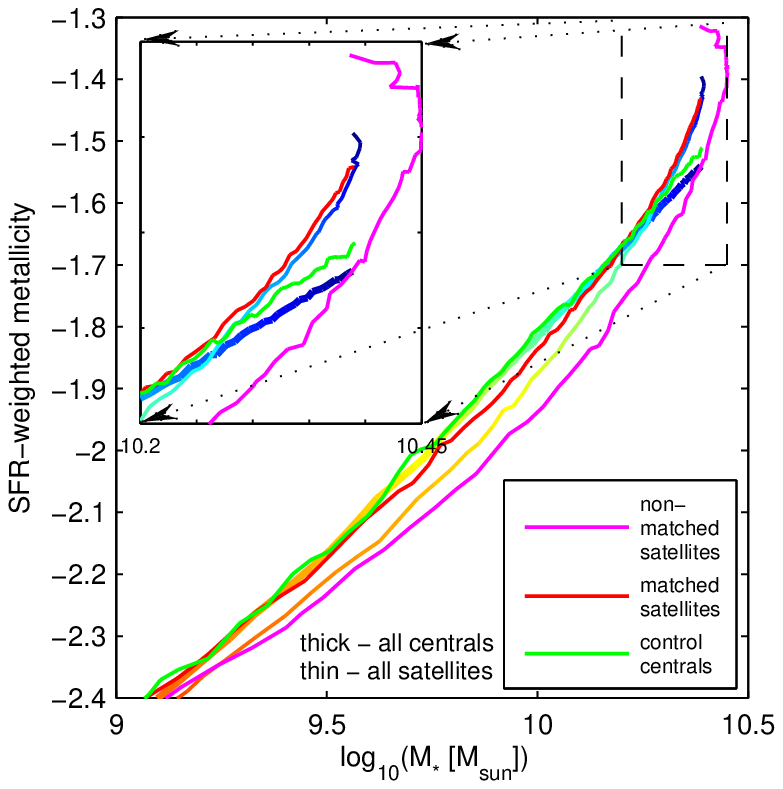}}
\caption{Evolution histories along the main-progenitor track in the mass-metallicity plane for various galaxy samples. The averaging of mass and metallicity is performed as a function of redshift. The insets zoom-in on the late-time evolution. Left: comparison between centrals (thick) and satellites (thin). The high-redshift main progenitors of satellites have a lower metallicity - at a given stellar mass - with respect to centrals, as they reach that same mass at an earlier time. However, toward $z=0$, corresponding approximately to the infall time, the trend reverses, and satellites increase sharply in metallicity, to have higher metallicities than centrals. Right: the central (thick rainbow) and satellite (thin rainbow) samples are repeated from the left panel. In addition, the satellite sample is split to the `matched' (red) and `non-matched' (magenta) samples, and the `control centrals' are also added (green). The `matched' satellites (red) and their `control' centrals (green) are by construction closely matched in terms of growth history. Here, they appear as closely related in terms of metallicity history as well, until late times, when the satellites show a sharp increase in metallicity.}
\vspace{0.3cm}
\label{f:MZhistory}
\end{figure*}

So far, we have established that satellite galaxies acquire their higher metallicities after, not before, becoming satellites. A second important conclusion from \Fig{samples_Z} can be drawn by dividing the effect shown in \Fig{MZrelation}, namely, that satellites tend to have higher metallicities, into two contributions. First, part of this is driven by the `non-matched satellites' sample, which has low $z=0$ SFRs, and hence -- by virtue of the anti-correlation between metallicity and SFR -- also higher metallicities. However, in addition, {\it despite being matched in SFR}, the `matched satellites' have higher metallicities than the `control centrals' by $\approx0.05-0.1\dex$.

To see this in more detail, \Fig{MZhistory} presents the evolution history of the various samples on the metallicity-stellar mass plane. In the left panel, the mean redshift of the track is indicated by the color, and only two samples are shown: all centrals (thick) and all satellites (thin). The inset shows a blow-up of the late-time evolution where, at $z\sim0.5$, the metallicity of the satellites shoots up with little mass evolution, representing the late-time increase we observed in \Fig{history_tinfall}. \footnote{For most of cosmic time, before the typical infall time of $z\sim0.5$, central galaxies have a higher metallicity than satellites at a given mass. While this trend may seem like the opposite of that discussed so far, one should keep in mind that a comparison at a given stellar mass is a comparison at different cosmological times, for centrals versus satellites. When considering the full central and satellite populations rather than the formation-matched `matched satellites' and `control centrals' samples, the satellites on average form earlier than centrals. Hence, at a given mass, satellites are found at an earlier epoch, and hence have lower metallicities.} The right panel of \Fig{MZhistory} presents the evolution of all of the samples we have discussed so far on the metallicity-stellar mass plane. Most notably, the `non-matched satellites' (magenta) have a very distinct evolution history. They form earlier, and hence have the lowest metallicity for a given stellar mass for most of the evolution track, but at late times they experience a sharp increase in metallicity with respect to mass. In fact, on average, they even lose mass at late times, while continuing to grow in metallicity. However, since they represent the minority of the total satellite population, the latter (thin rainbow curve repeated from the left panel) is much closer to the evolution of the `matched satellites' (red). Among the central galaxies, the full central population (thick rainbow curve repeated from the left panel) has a very similar evolution history to the `control centrals' (green). The latter is only shifted slightly ($\approx0.02\dex$) toward higher metallicities at late times. This can be attributed to the fact that the `control centrals' are slightly biased with respect to all centrals in terms of their formation history, i.e., they have slightly higher past SFRs and lower $z=0$ SFRs, since they are chosen to match the satellite galaxies in these parameters.

All in all, about $1/3$ of the difference at $z=0$ (the end of each curve in the right panel of \Fig{MZhistory}) between the full satellite (thin) and central (thick) populations can be accounted for by the `selection effect' on formation histories (the difference between the thin rainbow and red, plus the difference between the thick rainbow and green). Most of the difference, however, can be accounted for by the difference between the two sub-populations that {\it are matched} in terms of formation histories (green and red). That is exactly the difference that occurs after infall seen directly in \Fig{Zhistory_tinfall}.

\subsection{Spatially resolved and radially averaged metallicities}
\label{s:resolvedZ}

\begin{figure*}
\centering
\subfigure[]{
          \label{f:profileZcumulative}
          \includegraphics[width=0.49\textwidth]{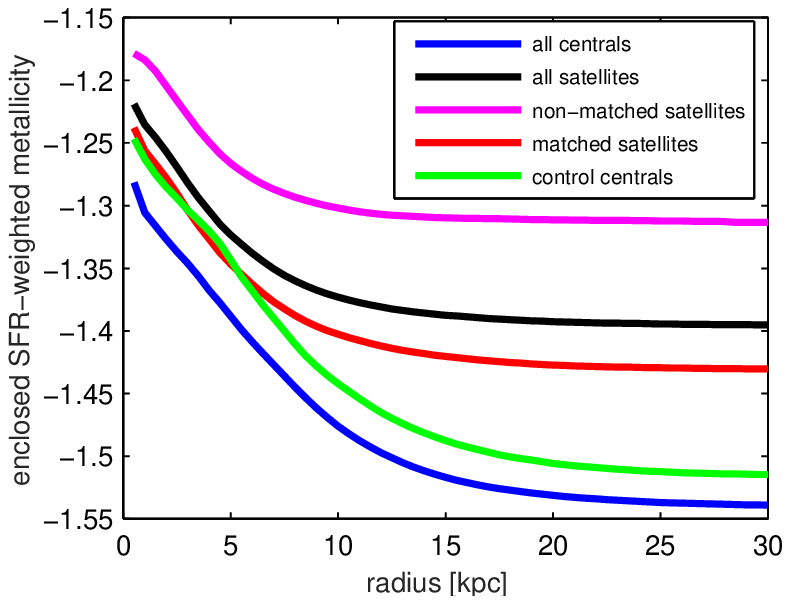}}
\subfigure[]{
          \label{f:profileZlocal}
          \includegraphics[width=0.49\textwidth]{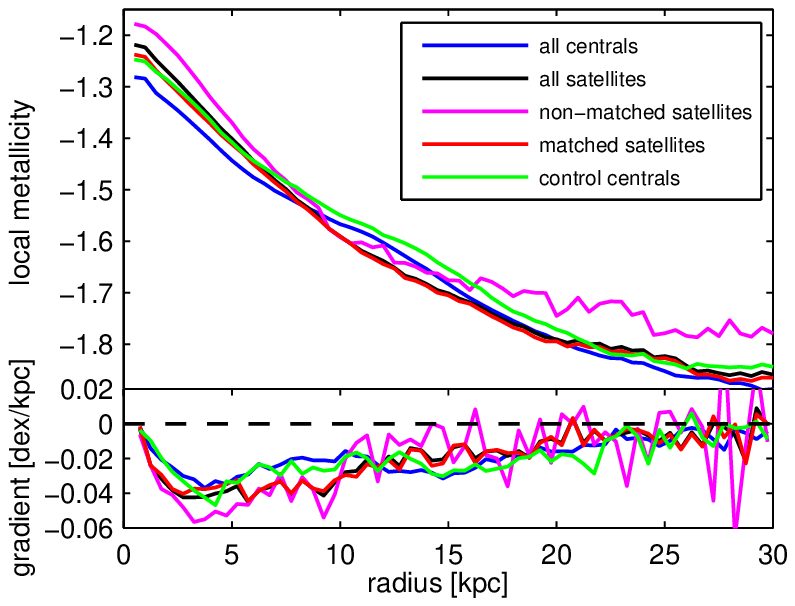}}
\caption{Left: the mean enclosed SFR-weighted metallicity of various $z=0$ galaxy samples, as a function of galactocentric radius. The largest radius of $30\kpc$ encloses essentially the full extent of the galaxies, hence the metallicity values there correspond to the $z=0$ values seen in \Fig{MZhistory_samples}. The $\approx0.1\dex$ difference between the `matched satellites' (red) and the `control centrals' (green) is seen here to only develop at radii larger than $\gtrsim7\kpc$, a radius inside which those two samples have essentially identical metallicities. Right, top: the mean local metallicity of various $z=0$ galaxy samples. The `matched satellites' (red) and `control centrals' (green) samples are almost indistinguishable at most radii, and where they do differ ($\approx10-20\kpc$), it is in the opposite sense to that seen in the left panel for the enclosed metallicity. This shows that it is not local metallicity differences that are responsible for the difference in enclosed metallicity, but instead a different radial weighting. Right, bottom: the derivative of the quantity in the top panel, i.e.~the metallicity gradient. The values of $\approx0.01-0.05\dex\kpc^{-1}$ as well as the tendency towards smaller gradients at larger radii are in good agreement with observations \citep{HoI_15a,Sanchez-MenguianoL_16a}.}
\vspace{0.3cm}
\label{f:profileZ}
\end{figure*}

In the previous section, we saw that satellites develop a higher metallicity than centrals after their infall into their host halo, and that some of this can be accounted for by their different SFR histories with respect to the centrals. However, most satellites can be matched in terms of formation history to a central galaxy, and even those matched samples differ significantly in $z=0$ metallicity, a difference that constitutes at least half of the total difference between the general satellite and central populations. In this section, we explore what accounts for the different metallicities of these formation-matched samples by spatially resolving them and looking into their structures.

\Fig{profileZcumulative} shows the mean enclosed SFR-weighted metallicity of the different samples as a function of galactocentric radius. For all samples, the enclosed metallicity is highest when only small radii are considered, it drops gradually when larger radii are taken into account in the averaging, and finally approaches a plateau that is equivalent to the galaxy-wide average, around the radius where the star-forming disk is diminished and there are no additional contributions of star-forming gas. As seen in \Fig{MZhistory}, there are slight shifts between `all centrals' (blue) and `control centrals' (green), as well as between `all satellites' (black) and `matched satellites' (red), which together (at the maximum radius shown, i.e.~the full galaxy-wide value) constitute $\approx40\%$ of the total difference between `all satellites' (black) and `all centrals' (blue). Importantly, these formation-history-driven differences are quite consistent with radius. However, the difference between the formation-history-matched samples, the `matched satellites' (red) and their `control centrals' (green), is strongly radius-dependent. In fact, at small radii, when looking only inside $\approx7\kpc$, those two samples of satellites and centrals have indistinguishable metallicities. Around that radius, the `matched satellites' sample starts plateauing, indicating that that is roughly the extent of the star-forming disks in that sample. In contrast, the enclosed SFR-weighted profile of the `control centrals' continues to drop, indicating that there are still sufficiently substantial amounts of star-forming gas at $\gtrsim7\kpc$ with low enough metallicity to drive the cumulative enclosed value down further.

\Fig{profileZlocal} shows, rather than the enclosed metallicity, the local metallicity profile as a function of radius. Similar to the case for the enclosed metallicities, the differences between `all centrals' (blue) and `control centrals' (green), as well as between `all satellites' (black) and `matched satellites' (red), are approximately radius-independent and, in fact, essentially fully explain the differences in the enclosed metallicities seen in \Fig{profileZcumulative}. However, the difference between the `matched satellites' (red) and their `control centrals' (green) is very small at all radii, as opposed to the case for the enclosed metallicities. In fact, in the range $\approx10-20\kpc$, the `control centrals' even have higher metallicities than the `matched satellites'. This clearly demonstrates that the higher {\it enclosed} metallicities that `matched satellites' have (at large radii) are not a result of differences in the {\it local} metallicity profile. Instead, they are a result of the satellites' `truncated' star-forming disks, as the enclosed SFR-weighted measurement samples mostly their inner, more metal-rich parts. This is in contrast with the `control centrals', which have the same local metallicity profile as the `matched satellites' (or even somewhat higher), but for which the SFR-weighted metallicity also includes substantial contributions from the large-radii, more metal-poor parts.

\begin{figure}
\centering
\includegraphics[width=0.475\textwidth]{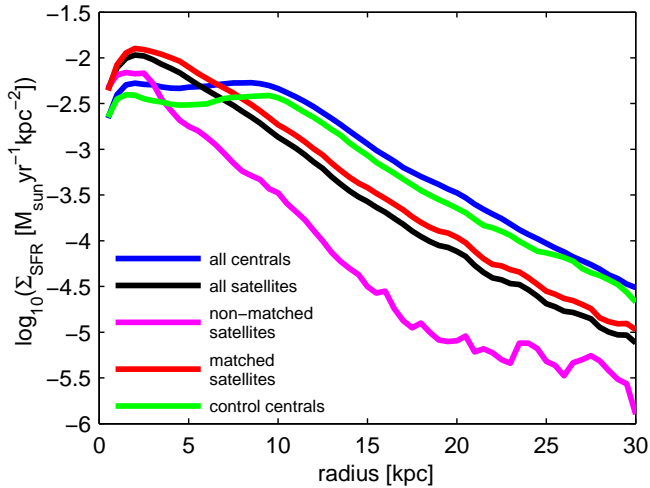}
\caption{Mean local SFR surface density of various $z=0$ galaxy samples as a function of galactocentric radius. All samples show an approximate exponential profiles at large radii ($\gtrsim10\kpc$), with central galaxies having mean surface densities $\approx0.3\dex$ higher than satellites (except for the `non-matched' satellites, which are further suppressed). Central galaxies, however, have a flat mean profile within $\approx10\kpc$, while the exponential profile of satellite galaxies holds down to $\gtrsim2\kpc$, giving them higher surface densities in the central parts. Hence, satellite galaxies have less extended, more concentrated, star-forming disks, than central galaxies.}
\vspace{0.3cm}
\label{f:profileSFR}
\end{figure}

The inevitable conclusion from \Fig{profileZ} regarding the differences in SFR profiles, namely, that satellites have more centrally concentrated star formation, is shown directly in \Fig{profileSFR}, which presents the mean profiles of SFR surface density for the various samples. It can be seen that central galaxies have nearly flat SFR surface densities in their inner $\approx10\kpc$, and outward of that they show a nearly exponential profile. In contrast, satellite galaxies have roughly exponential profiles all the way down to $\approx2\kpc$, with a lower normalization in the outer parts, but reaching higher surface densities than centrals at $\lesssim7\kpc$. This difference in profile shapes, together with the fact that metallicity gradients are generally negative as seen in \Fig{profileZlocal}, `biases' the overall SFR-weighted metallicity values of satellite galaxies toward higher values with respect to centrals, even though locally as a function of radius they have the same metallicity profiles.

Another interesting point to note regarding \Fig{profileSFR} and its relation to \Fig{profileZlocal} is the anti-correlation between the profile differences of the local metallicity and SFR surface density. In \Fig{profileSFR}, it can be seen that the subset of satellites that are `matched' (red) have somewhat higher SFR surface densities -- in a roughly radius-independent way -- than the total satellite population (black).\footnote{A difference that is balanced by the significantly lower SFR surface densities of the `non-matched satellite' sample (magenta).} \Fig{profileZlocal} shows that the opposite is true for the metallicity, i.e.~the metallicity of the `matched' sample is slightly {\it lower} than that of the whole satellite population. Similarly, for the centrals, the subset of `control' centrals (green) have {\it lower} SFR surface densities than `all centrals' (blue), with a very similar profile shape, but in turn {\it higher} metallicities. This demonstrates again, now in a spatially resolved fashion, that SFR differences drive some of the metallicity differences between the different samples. We note that while some work emphasizes the more fundamental role of the gas fraction over the SFR (e.g.~\citealp{ZahidJ_14a}), in our simulation, those are essentially the same due to the implementation of the star formation law. We checked explicitly and verified that the conclusions throughout this work would be unchanged if the star-forming gas mass were used instead of SFR.

\begin{figure*}
\centering
\subfigure[]{
          \label{f:profileSFR_evolution_sat}
          \includegraphics[width=0.32\textwidth]{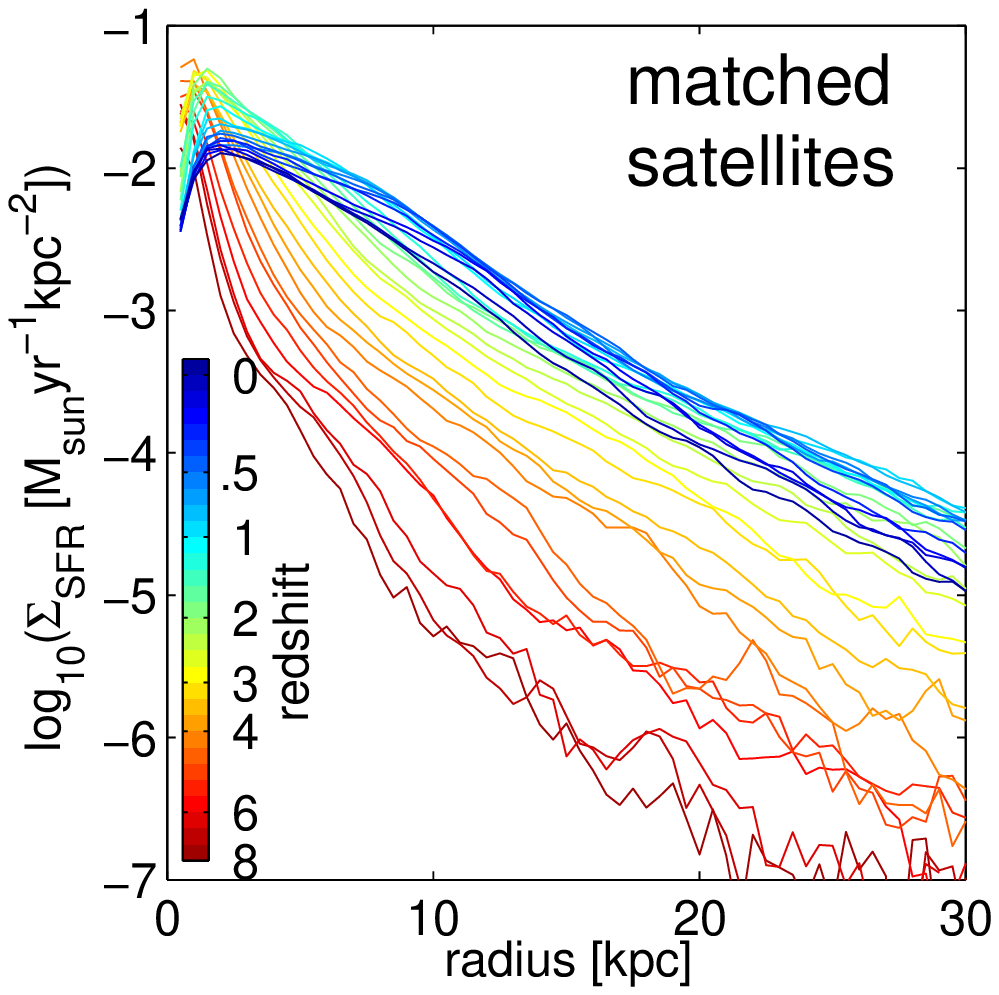}}
\subfigure[]{
          \label{f:profileSFR_evolution_cent}
          \includegraphics[width=0.32\textwidth]{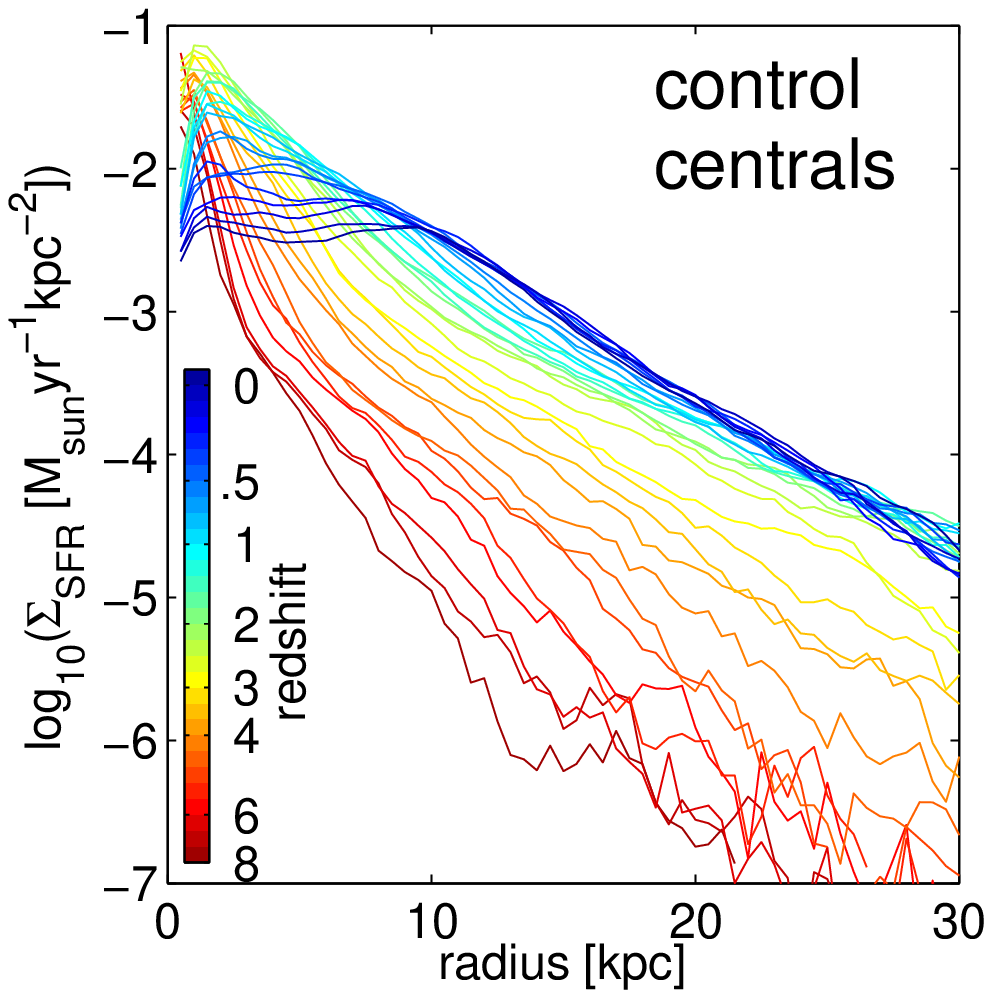}}
\subfigure[]{
          \label{f:profileSFR_evolution_diff}
          \includegraphics[width=0.32\textwidth]{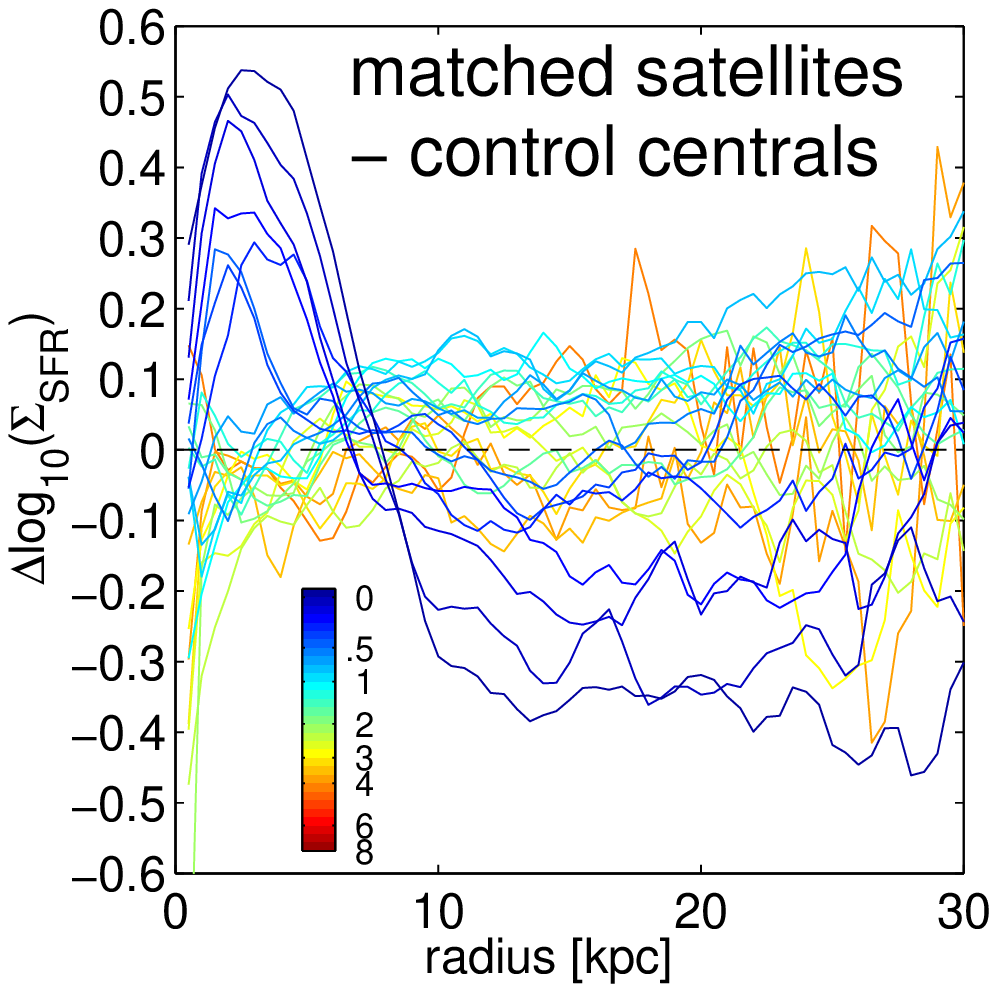}}
\caption{Evolution over cosmic time of the mean SFR surface density for the `matched satellite' sample (left), their matched `control central' sample (middle), and the difference between the two (right). The two samples show very similar SFR profiles until the typical infall epoch of $z\sim0.5-1$. Thereafter, in spite of being matched in formation histories down to $z=0$, the large-radius part of the mean SFR profile decreases with time for the satellite galaxies, but remains approximately constant for the central galaxies. At the same time, the inner part evolution shows opposite trends, with central galaxies developing a flat profile that diminishes with time. As a result, satellite galaxies have higher inner SFR surface densities than central galaxies, and lower outer densities.}
\vspace{0.3cm}
\label{f:profileSFR_evolution}
\end{figure*}

Before moving on with our discussion of metallicities, we briefly look in more detail at the SFR surface density profiles seen in \Fig{profileSFR}. To this end, \Fig{profileSFR_evolution} presents the evolution over cosmic time of the mean SFR surface density profiles along the main-progenitor tracks of two samples, the `matched satellites' (left) and `control centrals' (middle), as well as the difference between the two (right). As can be seen in the right panel, the two populations, which are matched in their formation histories (see \Figs{samples}{Mhistory_tinfall}), have essentially the same SFR surface density profiles until $z\approx2$. Between $z\approx2$ and $z\approx1$, the satellites have slightly more extended profiles, with lower SFR surface densities in the inner parts and higher in the outer parts. However, at $z\approx0.5$ -- the typical infall time, as seen in \Fig{history_tinfall} -- the picture reverses, i.e.~satellites develop a more concentrated SFR profile, and the differences become significantly larger. After $z\approx3$, the mean inner profiles drop over time for both samples, but after infall time, $z\approx0.5$, the drop for the satellite sample is halted, while it continues for the centrals, resulting in higher inner profiles for satellites than centrals at $z=0$. The outer profiles evolve differently: for the centrals, they constantly increas with cosmic time until they remain approximately constant at $z\lesssim1$. On the other hand, the outer profiles of satellites reach a peak at $z\approx1$ and thereafter decrease with time. This results in satellites that have lower outer SFR surface densities than centrals at $z=0$. It is worth noting that while the overall normalizations change somewhat if we replace the `matched satellites' sample with the full satellites sample, and/or the `control centrals' with the full centrals samples, the differential effect between the inner and outer parts remains essentially unchanged.

While a detailed explanation of these trends is beyond the scope of this paper, we note that the outer post-infall drop of the satellite SFR surface densities is likely associated with stripping events \citep{BaheY_15a}. It may also originate from outside-in quenching resulting from strangulation. The evolution of the inner parts is perhaps more puzzling and merits future investigation. One possibility is that the interstellar gas in the inner parts of satellites is compressed due to the additional external pressure provided by the host halo atmosphere, promoting star formation. Another possibility is that the inner parts of satellites are experiencing elevated SFRs due to interactions and stronger tidal forces. It is also worth noting that preliminary observational evidence exists that satellite galaxies indeed show more concentrated star formation relative to centrals \citep{BrethertonC_13a}. Larger observational samples and further analysis are required to draw firm conclusions concerning the evolution of star-forming disks in rich environments. These will provide important constraints for cosmological hydrodynamical simulations such as Illustris and, in fact, may also have the power to tell whether the scenario we describe here to explain satellite galaxy metallicities based on the Illustris simulation is also valid in reality.

\begin{figure*}
\centering
\includegraphics[width=1.0\textwidth]{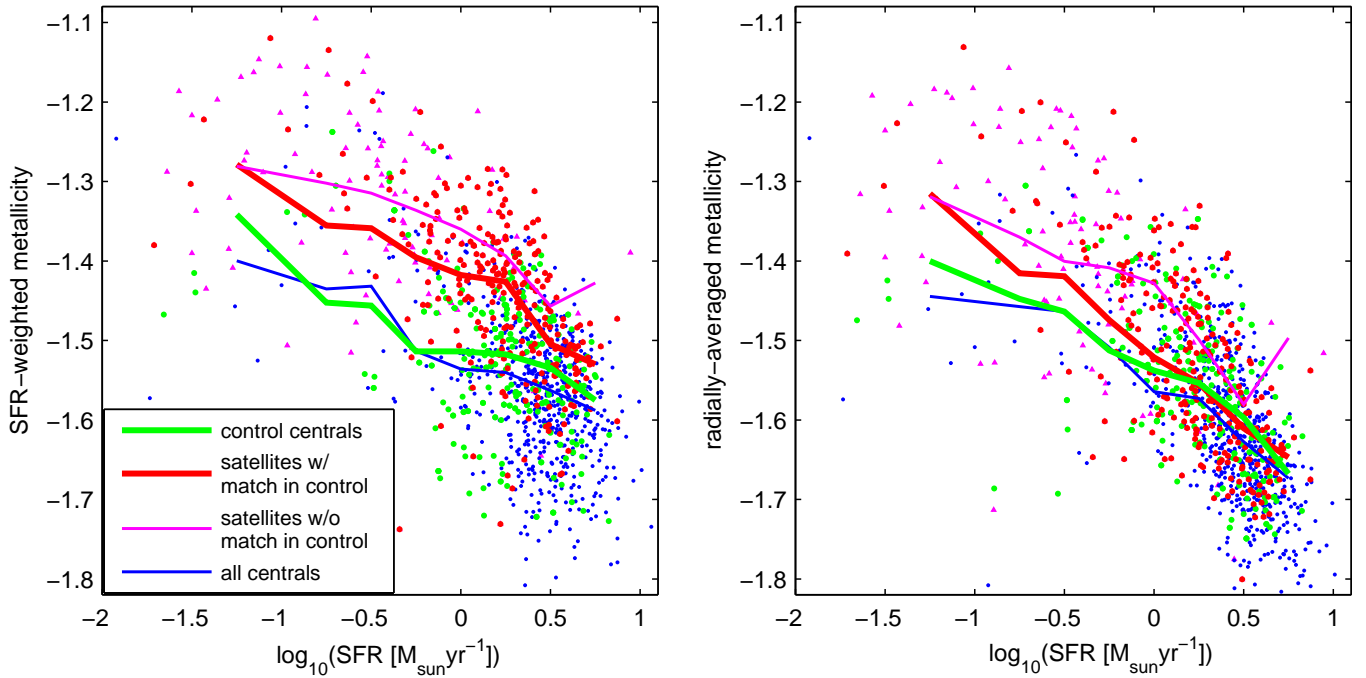}
\caption{The $z=0$ relation between metallicity and SFR for various galaxy samples, showing both individual galaxies (points) and running medians (curves), with two weighting schemes: SFR-weighted (left) and radially averaged (right). A clear anti-correlation between metallicity and SFR is seen for all samples. The relation is almost identical for the vast majority of galaxies ($\approx80\%$ of the galaxies have $\SFR>1\Msunyr$) when the metallicity is radially averaged (right), i.e.~when the local metallicity is considered. This is true in particular for the two samples that are matched in formation history, i.e.~the `matched satellites' (red) and their `control centrals' (green). However, when SFR-weighted metallicities are considered, satellites tend to have higher metallicities than centrals, even at a given SFR and for a matched formation history (left).}
\vspace{0.3cm}
\label{f:Z_SFR_definitions}
\end{figure*}

\begin{figure}
\centering
\includegraphics[width=0.475\textwidth]{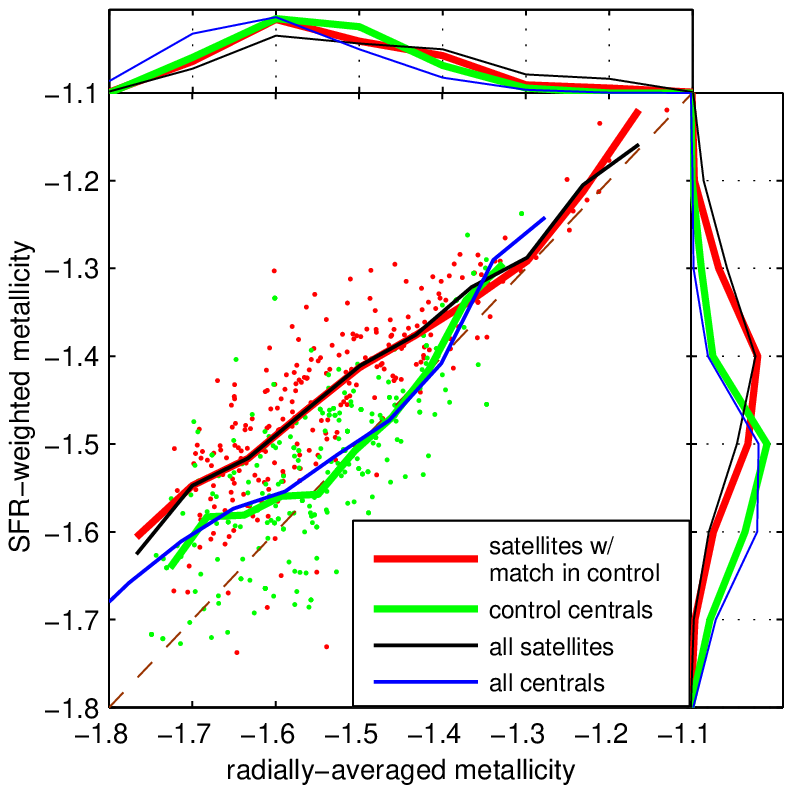}
\caption{Direct comparison, for various galaxy samples, of the two metallicity definitions considered in this work: SFR-weighted metallicity on the y-axis versus radially averaged metallicity on the x-axis. Curves indicate running medians, and the dashed curve represents the identity relation. Central galaxies (blue, green) usually have very similar radially averaged and SFR-weighted metallicities (except at low metallicities). Satellite galaxies (black, red) have a different relation, with the SFR-weighted metallicity typically $\sim0.1\dex$ higher than the radially averaged metallicity. The top and right panels show the projected distributions of these two quantities.}
\vspace{0.3cm}
\label{f:definitions_comparison}
\end{figure}

Returning to the metallicities, the difference in the enclosed metallicities between the `matched satellites' and `control centrals' samples (\Fig{profileZcumulative}), which emerges once their essentially equivalent local metallicity profiles (\Fig{profileZlocal}) are weighted by their SFRs, motivates us to introduce a new quantity that represents the local metallicity profile. We therefore simply average the local metallicity profiles shown in \Fig{profileZlocal} in a way that gives equal weight to each radius, and we refer to this quantity as the `{\bf radially averaged metallicity}'. Note that it is not weighted by SFR, neither by mass, volume, or area. In \Fig{Z_SFR_definitions}, we present the $z=0$ distributions of metallicity and SFR for the various samples, using the two metallicity definitions: SFR-weighted (left) and the newly defined `radially averaged' (right). The left panel repeats the right panel of \Fig{samples_Z} but with the addition of central galaxies. The separation previously noted in the discussion of \Fig{samples_Z} between the `matched satellites' and their `control centrals' is seen here more explicitly with the introduction of running medians (curves). In contrast, the right panel shows no difference in the radially averaged metallicity between the two formation-matched samples (with the exception of the low-SFR, $<1\Msunyr$, regime where only $20\%$ of these galaxies lie). \Fig{definitions_comparison} shows a direct comparison between the two metallicity definitions. For central galaxies, the SFR-weighted metallicity tends to be equal to the radially averaged metallicity. Satellite galaxies, however, occupy a different locus on this plane, having SFR-weighted metallicities typically $\approx0.1\dex$ higher than their radially averaged metallicities. This difference is the result of the different weighting schemes where the SFR-weighted metallicities weigh the inner regions more strongly; this effect is more pronounced for satellite galaxies.

\begin{figure}
\centering
\includegraphics[width=0.475\textwidth]{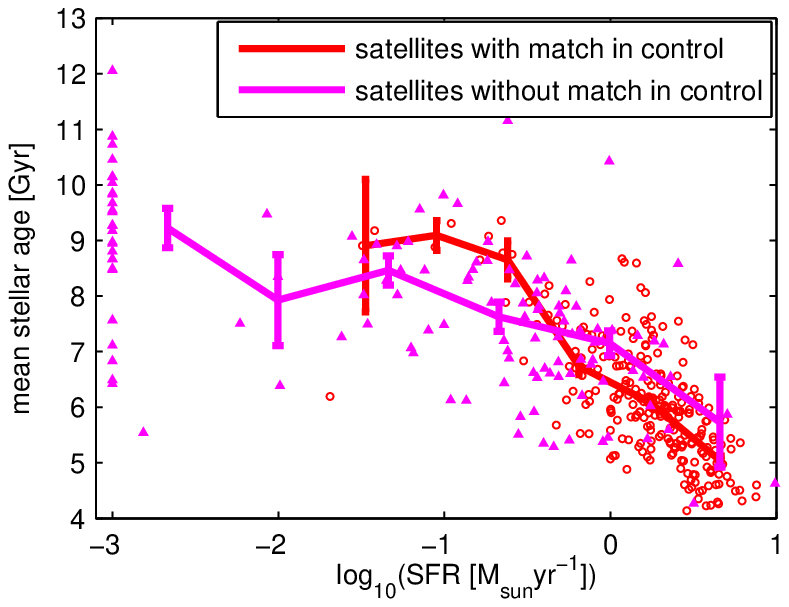}
\caption{Mean stellar age of the stellar populations inside twice the stellar half-mass radius (corresponding roughly to the size of the star-forming disk) versus the $z=0$ total galaxy SFR, showing individual galaxies (points) as well as running medians (curves). Galaxies with higher $z=0$ SFR tend to also be younger. For $\SFR\gtrsim0.5\Msunyr$, galaxies belonging to the `non-matched' satellite sample (magenta) tend to have older stellar populations than galaxies belonging to the `matched satellites' sample (red), {\it at a given SFR}. This means that even at a given $z=0$ SFR, `non-matched' satellites tend to form earlier than `matched' satellites. The galaxies shown at $\SFR=10^{-3}\Msunyr$ represent galaxies that in fact have $\SFR=0$.}
\vspace{0.3cm}
\label{f:stellarage_SFR}
\end{figure}

It is interesting to note that even for the radially averaged metallicity, there remain small differences in the median relations between the other two samples, which are not formation-matched. In particular, while in \Fig{profileZlocal} we have seen that `non-matched satellites' have higher radially averaged metallicities than `matched satellites', in the right panel of \Fig{Z_SFR_definitions} we see that this holds even {\it for a given SFR}. This means that even when the different weighting is taken away by using the radially averaged metallicities, considering only the instantaneous SFR does not fully account for the metallicity differences. We argue that this is very plausibly a result of the different formation histories. In \Fig{stellarage_SFR}, we show that even at the same $z=0$ SFR, `non-matched' satellites tend to have older stellar ages (at $\SFR\gtrsim0.5\Msunyr$, where most galaxies lie, and where most of the difference between the two populations in \Fig{Z_SFR_definitions} appears). This can account for their higher metallicities compared to the `matched' satellites as a result of two distinct physical effects. First, their older stellar populations had more time to release metals back into the gas. Second, their younger stellar ages imply earlier formation histories, i.e.~most of their wind-launching occurred at higher redshift. Since the wind mass-loading factors in our model are lower at higher redshifts (at a given stellar mass), galaxies with earlier formation histories (at a given final stellar mass) are likely to have ejected less of their metals out of the galaxy, resulting in higher $z=0$ galaxy metallicities. Another possibility, not directly related to the formation history but which is motivated by the fact that galaxies in the `non-matched' sample have spent more time as satellites than the `matched' ones (not shown) and have more strongly truncated SFR profiles (\Fig{profileSFR}), is that the environmental metallicity does play some role for this sample (see the discussion of environmental metallicities in the next section).

\section{Main results: environmental overdensity}
\label{s:envir}
Thus far, we have only considered the binary distinction between central and satellite galaxies. In this section, we consider the dependence of metallicity on a continuous measure of environment, namely, the environmental overdensity, defined in Section \ref{s:methods}. We follow \citet{PengY_14a}, who recently quantified galaxy metallicity as a function of environmental overdensity in SDSS, importantly, while separating between central and satellite galaxies. \citet{PengY_14a} found that while satellite galaxies exhibit a positive correlation between metallicity and environmental overdensity, central galaxies hardly ever do. They interpreted this finding as a result of more metal-rich accretion at denser environments. In this section, we show that similar trends exist in Illustris, but we find that their origin is different than that hypothesized in \citet{PengY_14a}.

\begin{figure*}
\centering
\subfigure[]{
          \label{f:Z_env_3kpc}
          \includegraphics[width=0.32\textwidth]{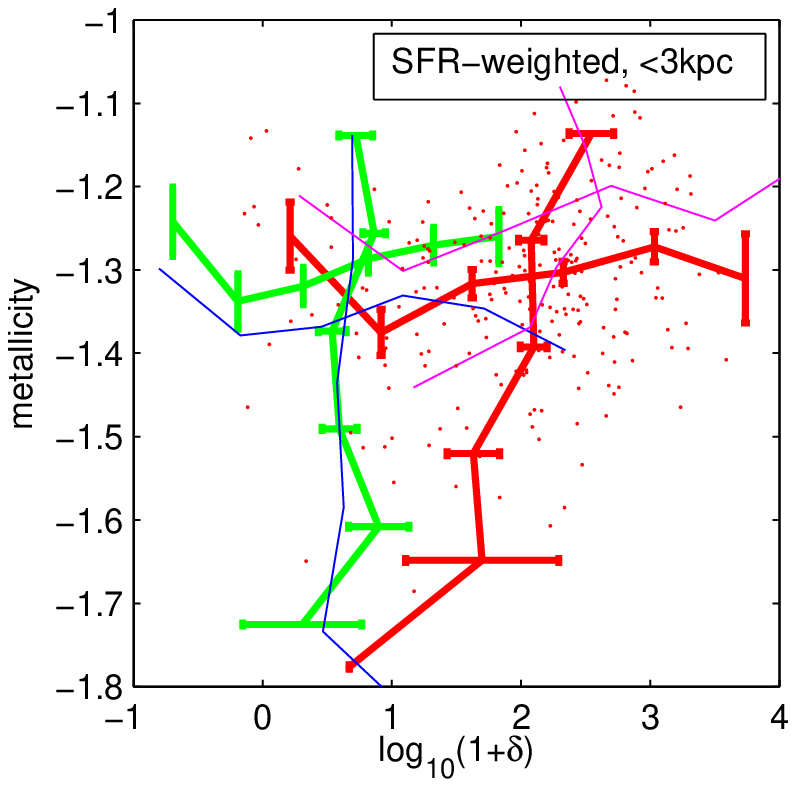}}
\subfigure[]{
          \label{f:Z_env_30kpc}
          \includegraphics[width=0.32\textwidth]{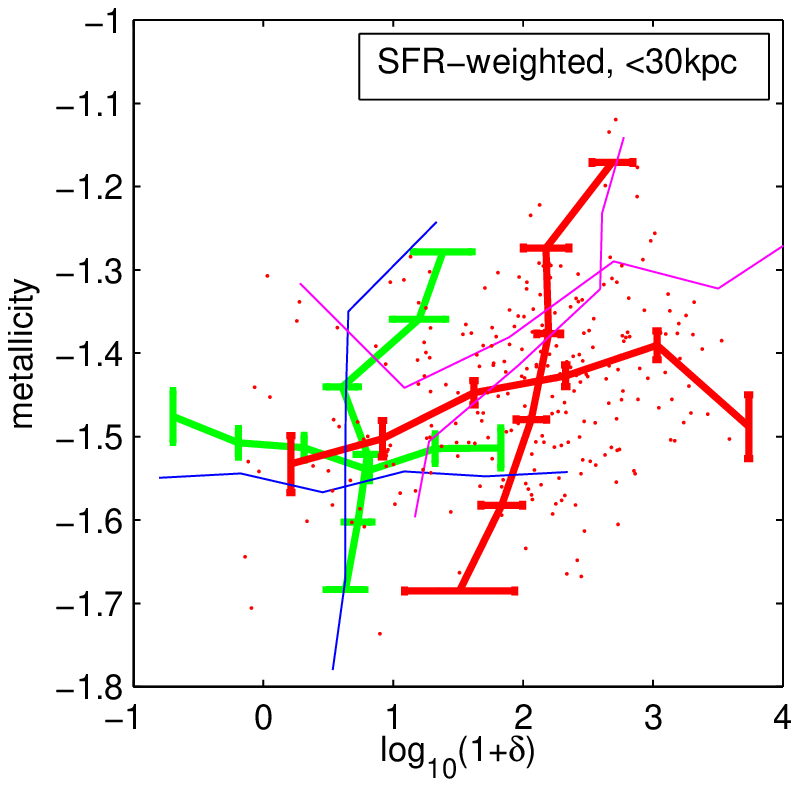}}
\subfigure[]{
          \label{f:Z_env_radavg}
          \includegraphics[width=0.32\textwidth]{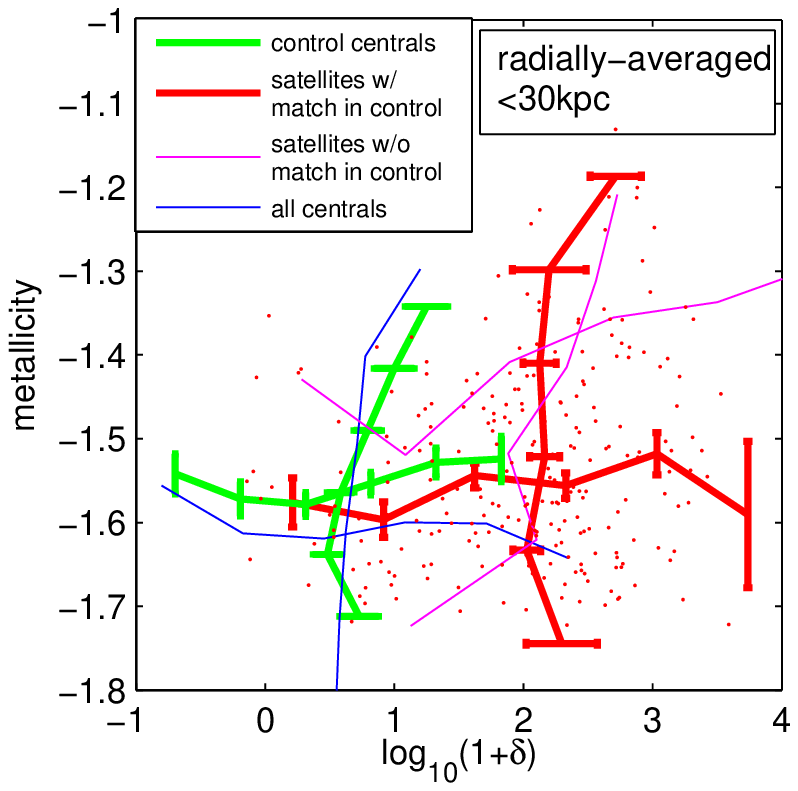}}
\caption{Various galaxy samples on the metallicity-environment plane, for different definitions of metallicity. The two-dimensional distributions are represented by running medians: the approximately horizontal curves show median metallicity versus environmental overdensity, and the approximately vertical curves show median environmental overdensity as a function of metallicity. Error bars represent uncertainties on the median values. Balancing between information content and readability, the full distribution is also shown, but only for the `matched satellites' (red points). Central galaxies (all centrals: blue, control centrals: green) show essentially no correlation of metallicity with environment, in agreement with observations \citep{PengY_14a}. `Matched satellites' (red), i.e.~those with a formation history that can be matched by a central galaxy, show a clear correlation when SFR-weighted metallicities are considered at full radius (middle panel), but a weaker correlation at smaller radii (left panel). When the radially averaged metallicity is considered (right panel), there is essentially no correlation. Hence, the correlation with environment displayed by satellites is a result of formation history and stripping, not of environmental metallicity.}
\vspace{0.3cm}
\label{f:Z_env}
\end{figure*}

\Fig{Z_env} presents the two-dimensional plane of metallicity and environmental overdensity, using three different measures of metallicity. In \Fig{Z_env_3kpc}, the metallicity is SFR-weighted, but only inside the inner $3\kpc$ of each galaxy. In \Fig{Z_env_30kpc}, the metallicity is SFR-weighted within $30\kpc$, which is essentially also the galaxy-wide SFR-weighted metallicity. In \Fig{Z_env_radavg}, the metallicity is the radially averaged metallicity defined in Section \ref{s:resolvedZ}, which is not affected by the shape of the SFR profile. In each panel, the full distribution is shown only for the `matched satellites' sample (red points), and for visual clarity the distributions for the other samples are represented only by two curves each: the median metallicity as a function of overdensity (approximately horizontal curves) and the median overdensity as a function of metallicity (approximately vertical curves). Error bars, shown in the interest of visual clarity only for the `matched satellite' and `control central' samples, represent the standard error of the medians, defined as $1.253\sigma/\sqrt{N}$, where $\sigma$ is the standard deviation of the distribution of values in each bin, and $N$ is the number of galaxies included in each bin.

The first point we take away from \Fig{Z_env} is that central galaxies show no significant correlation between metallicity and environmental overdensity, as is evident from the lack of slope (full horizontality/verticality) in their curves (blue and green curves, for all centrals and `control centrals', respectively). This is in agreement with the observational results of \citet{PengY_14a}. For satellites, the picture is more subtle. We begin by considering the satellite population with `non-extreme' formation histories, i.e., those that can be matched to centrals (red). For these, a clear correlation is found between metallicity and environment, as is made evident by the positive slopes of the medians running in both dimensions. The strongest such correlation is found for the galaxy-wide SFR-weighted metallicities (\Fig{Z_env_30kpc}). A weaker, but still statistically significant, correlation is found for the $<3\kpc$ SFR-weighted metallicities (\Fig{Z_env_3kpc}). Once the SFR profiles are taken out of the equation, by examining the radially averaged metallicities, the correlation with environment becomes much weaker, perhaps even statistically insignificant (\Fig{Z_env_radavg}). In fact, the formation-matched samples are consistent with having the same relation between radially averaged metallicity and environment, as the horizontal red and green curves in \Fig{Z_env_radavg} overlap within the error bars. These two samples only occupy different regions in environmental overdensity space, but for a given environmental overdensity, `matched satellites' and `control centrals' seem to possess the same radially averaged metallicity. The interpretation of these trends is clear -- the correlation between metallicity and environment is driven by a correlation between SFR profile shape and environment, not by the local metallicity profile and environment. Since the `matched satellites' constitute most of the general satellite population ($\approx2/3$ of it), we conclude that a large part of the positive metallicity-environment correlation for the general satellite population (see \Fig{Peng}) is driven by this effect.

\begin{figure}
\centering
\includegraphics[width=0.475\textwidth]{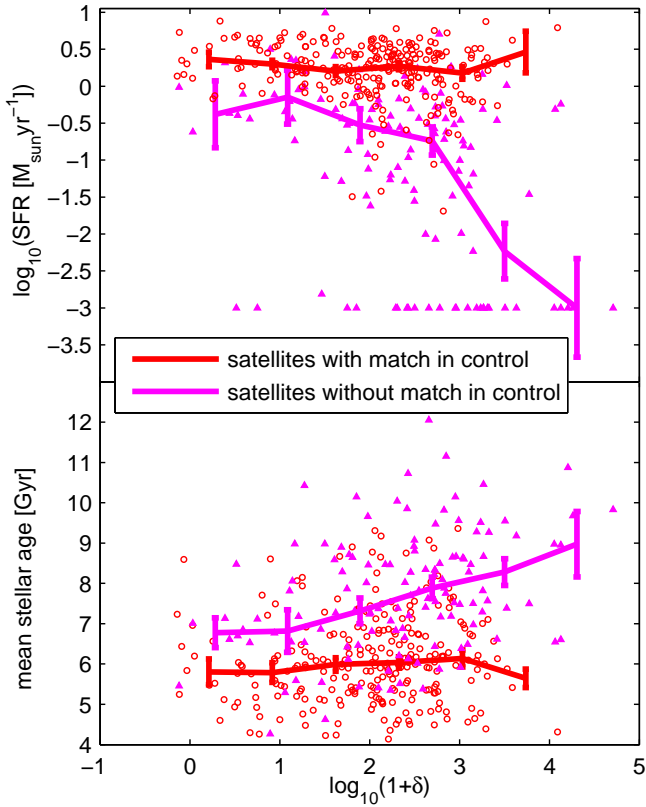}
\caption{The $z=0$ total SFR (top) and mean stellar age of the stellar populations inside twice the stellar half-mass radius (bottom) versus environmental overdensity, showing individual galaxies (points) as well as running medians (curves). For satellites with formation histories that can be matched to centrals (red) there are no significant correlations. Satellites that are not matched to centrals (magenta), on the other hand, show generally lower SFRs and older ages, i.e., earlier formation times, and also show significant correlations of those quantities with environmental overdensity. The galaxies in the top panel shown at $\SFR=10^{-3}\Msunyr$ represent galaxies that in fact have $\SFR=0$.}
\vspace{0.3cm}
\label{f:stellarageSFR_environment}
\end{figure}

\begin{figure*}
\centering
\includegraphics[width=1.0\textwidth]{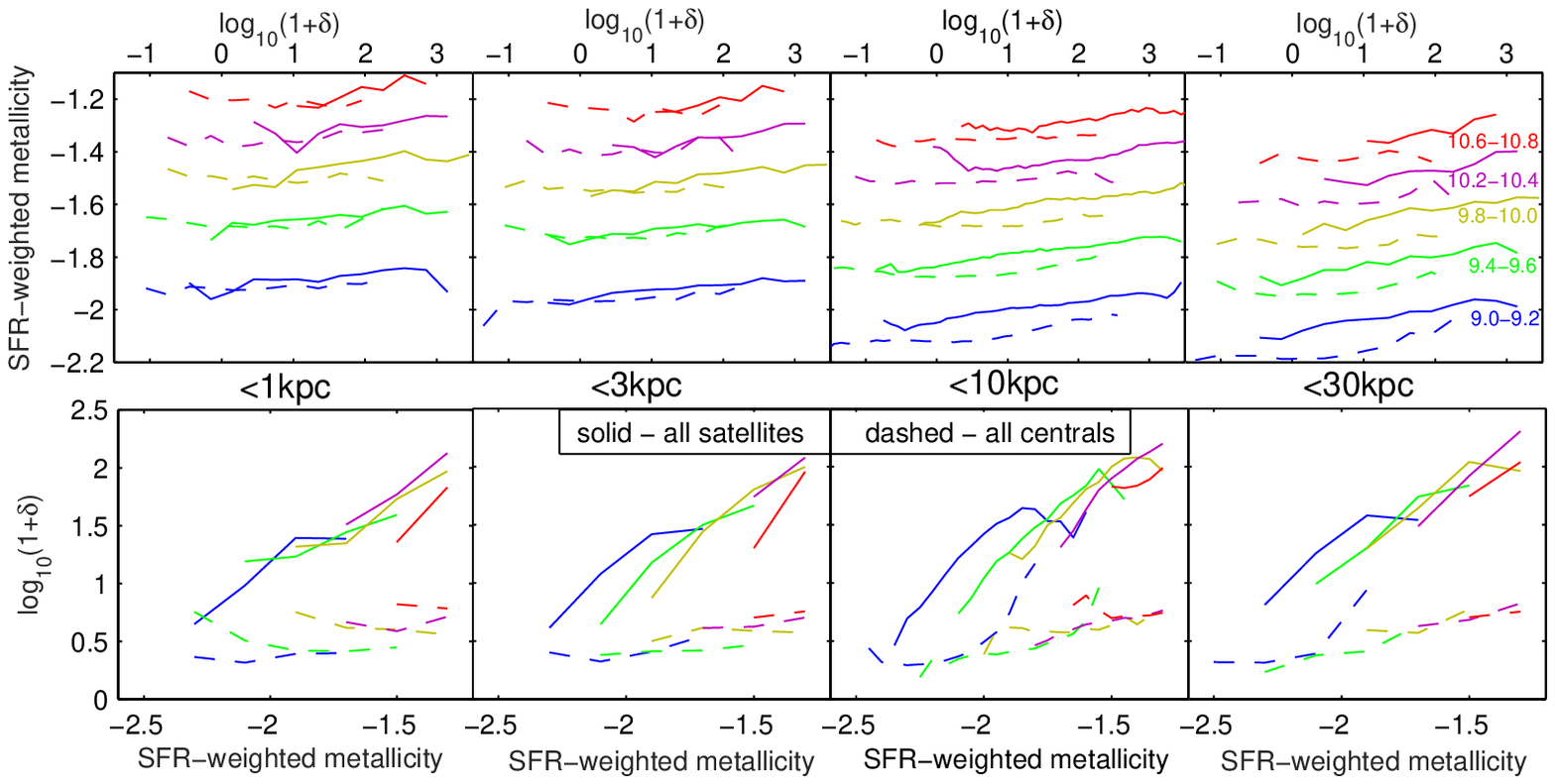}
\caption{Relations between SFR-weighted metallicity and environmental overdensity, for galaxies of various stellar masses (indicated in log units in the top-right panel), following the observational result shown in Figure (3) of \citet{PengY_14a}. Top: mean metallicity versus environment. When the full extent of galaxies is considered (right panel), satellite galaxies (solid) residing in denser environments tend to have higher metallicities. The metallicity of central galaxies (dashed) is significantly less sensitive to the environment. The smaller the `aperture' in which metallicity is measured (progressing from right to left panels), the less environment-dependent the metallicity becomes. Bottom: environment versus metallicity, displaying good qualitative agreement with \citet{PengY_14a}, in particular for small radii.}
\vspace{0.3cm}
\label{f:Peng}
\end{figure*}

However, a non-negligible fraction, namely, $\approx1/3$, of the general satellite population is composed of the `non-matched satellites'. These do present a positive metallicity-overdensity relation, even for the radially averaged metallicity (magenta curves, \Fig{Z_env_radavg}). To gain insight into this finding, in \Fig{stellarageSFR_environment} we plot the SFR (top) and mean stellar age (bottom) of the galaxies in the two satellite samples versus the environmental overdensity. The `matched satellites' show almost no correlation, while the `non-matched satellites' show significant strong correlations of lower SFR and larger mean stellar age in denser environments. These correlations provide several explanations for the positive correlation between metallicity and environment that this sample displays in \Fig{Z_env_radavg}, which are related to their formation history. First, they are `biased' to higher metallicities by virtue of the metallicity-SFR anti-correlation. In addition, as argued in the last paragraph of Section \ref{s:resolvedZ}, galaxies with larger mean stellar age, i.e., galaxies that formed earlier, are expected to have higher metallicities even at a given $z=0$ SFR due to the combination of a larger release of metal mass by their older stellar populations and a lower historical galactic winds-driven ejection of metal mass out of the galaxy.

The strong role, even if not the only role, that the concentration of the SFR profiles in satellite galaxies plays in their metallicity-overdensity relation has a simple observable implication, which is a prediction of our simulation. Namely, since the concentration effect of the SFR profiles is more pronounced the larger the considered aperture is, and since we find that a large part of the metallicity-overdensity correlation is driven by the SFR profile concentration effect, we predict that the metallicity-overdensity correlation is weaker when smaller apertures are considered. This was already noted in relation to \Figs{Z_env_3kpc}{Z_env_30kpc}. In \Fig{Peng} we show a more extensive comparison of different apertures in a format that can be directly compared to Figure 3 of \citet{PengY_14a}, and for several bins of stellar mass in particular, indicated by different colors. In the top panels, the mean SFR-weighted metallicities are shown against the environmental overdensity, for centrals (dashed) and satellites (solid) separately, and for different apertures ($1,3,10,30\kpc$ from left to right). The bottom panels show the mean overdensity as a function of metallicity. Indeed, the correlations are stronger for larger apertures. The metallicities presented by \citet{PengY_14a} are limited to the SDSS fiber, which for their redshift range corresponds to $\approx1.25-5\kpc$. However, the absolute scale of the apertures should not be compared at face value with observations, as Illustris galaxies are known to be generally larger than observed galaxies \citep{SnyderG_14a,WellonsS_16a}. The qualitative trends reported for SDSS galaxies by \citet{PengY_14a} are reproduced in \Fig{Peng}, which in turn also provides a direct prediction for the (at least qualitative) trend with aperture. Some correlation still exists even for very small apertures, as expected from \Fig{stellarageSFR_environment}, which shows an effect that is not directly aperture-related.

\begin{figure}
\centering
\includegraphics[width=0.475\textwidth]{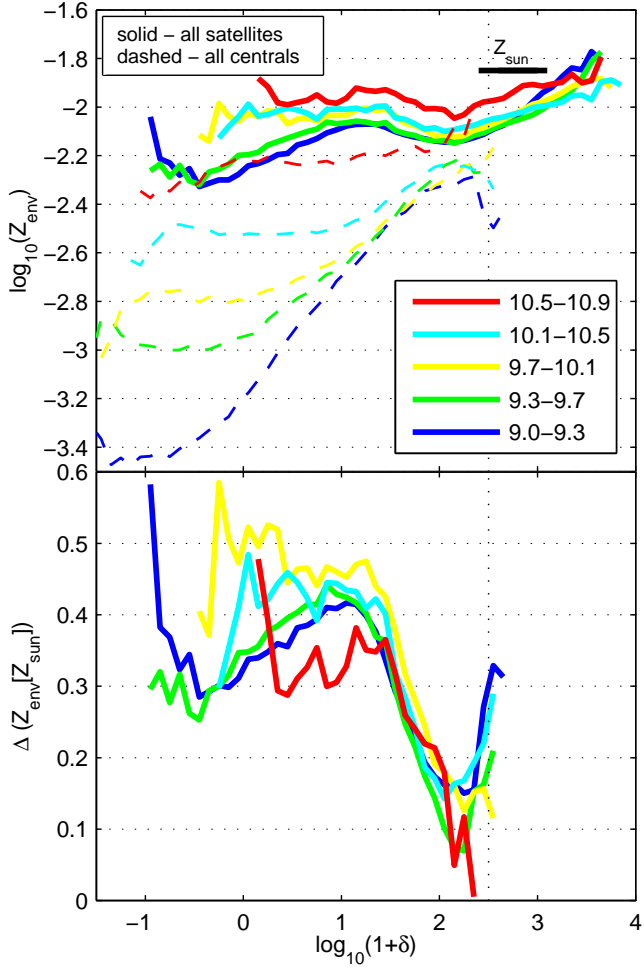}
\caption{Top: the mass-weighted metallicity of the inflowing circumgalactic (CGM) gas (see definition in the main text), as a function of environmental overdensity, for galaxies in different stellar mass bins (as indicated by the legend), separated by satellites (solid) and centrals (dashed). Satellite galaxies are seen to have their CGM enriched to $\approx0.5-1\Zsun$ with a weak dependence on mass and environment. Central galaxies have more enriched CGM at higher masses and denser environments, but generally have much less enriched CGM than satellites. Bottom: the difference, here in linear units of solar metallicity, between the CGM metallicity of satellites and centrals (the solid and dashed curves in the top panel, respectively), for different stellar mass bins, as a function of environmental overdensity. At low and medium overdensities ($\delta\lesssim10$), the difference is $0.3-0.5\Zsun$, and it becomes smaller at higher overdensities, mostly because the centrals galaxies have increased CGM metallicities.}
\vspace{0.3cm}
\label{f:environment}
\end{figure}

We have generally argued that the SFR-weighting effect and differences in formation histories account for most of the differences seen between the metallicities of satellites and centrals, and in particular also as a function of environmental overdensity. \citet{PengY_14a}, however, interpret their findings as a result of an environmentally dependent intergalactic metallicity that provides a metallicity floor for galaxies in different environments by means of gas accretion, see their Equation 2. In \Fig{environment} we are inspired by Figure (6) of \citet{PengY_14a} and plot the environmental metallicity, or circumgalactic metallicity, around galaxies in different mass bins, separated into centrals and satellites (top; in the logarithmic metallicity units used throughout the paper), as well as the difference between those circumgalactic metallicities between satellites and centrals (bottom; in linear units of solar metallicity). The circumgalactic metallicity is not a well-defined quantity, and hence its calculation is somewhat arbitrary. Here, we calculate it as the mass-weighted metallicity of all of the gas in a shell with inner and outer radii of factors of $10$ and $20$, respectively, times the stellar half-mass radius of each galaxy, which in addition has an inward radial velocity exceeding half of the maximum circular velocity of the galaxy. Since the stellar half-mass radii of Illustris galaxies are typically $\sim5\%$ of the virial radii of their dark matter halos, this definition selects gas that is in a shell mostly inside the halo but close to the virial radius. This is intended to be a simple selection of gas that is not directly related to the galaxy itself but is likely to be participating in the accretion onto the galaxy. We verified that our qualitative results do not change with modest modifications of the radii defining the shell or even by removing altogether the radial velocity criterion.

\Fig{environment} indeed shows that satellite galaxies have higher circumgalactic metallicities than central galaxies, both at a given stellar mass and environmental overdensity. The difference is $\approx40\%$ of the solar metallicity for low overdensities. It becomes smaller for higher overdensities, as the environments of the centrals also become more metal-rich, likely due to feedback from black holes \citep{SureshJ_14a}. Reassuringly, while the absolute circumgalactic metallicities depend on stellar mass (in particular for central galaxies where more massive ones are plausibly enriching their own environments to a larger degree), the difference between satellites and centrals is rather independent of stellar mass but instead depends more on environment. This is expected, since both centrals and satellites are expected to enrich their own environment to some (stellar mass-dependent) degree, but the difference between the two represents an effect that is external to the galaxies, and hence is more sensitive to environment.

The main point to be taken from \Fig{environment} is that the difference in circumgalactic metallicity between satellites and centrals, as measured here, is smaller but not negligible compared to the absolute metallicities of galaxies in the mass bin we have focused on in this work, $M_*=(2-3)\times10^{10}\Msun$. As can be seen in \Fig{MZrelation}, the metallicity of satellites in this mass bin is $\approx2.6\Zsun$, and that of centrals is $\approx2\Zsun$. The difference is somewhat larger but comparable to the difference in circumgalactic metallicities between satellites and centrals that we measure for galaxies in this mass bin for $\delta\lesssim10$, which is $0.4\Zsun$ (\Fig{environment}). Hence, it may be surprising that in \Fig{Z_env_radavg} we see no discernable difference between the radially averaged metallicities of satellites and centrals at these environmental overdensities once they have been matched in formation history. Perhaps the most plausible explanation is that the circumgalactic metallicity as defined here is not the most appropriate measure for the `metallicity of accretion' quantity that serves as a metallicity floor in analytical models, e.g.~Equation 2 in \citet{PengY_14a}. A more detailed analysis of gas accretion would be required to determine how to measure the most relevant quantity most accurately. However, the exercise presented here teaches us that even if the gas around the virial radius is enriched to significant levels (as also suggested by observations; \citealp{FaermanY_16a}), this is not necessarily manifested in the metallicities inside galaxies once the formation history and weighting effects have been accounted for.

\section{Summary and conclusion}
\label{s:summary}
In this paper, we present a study of the relations between galaxy gas-phase metallicity and environment in the Illustris simulation. The basic result, and the starting point for our study, is that satellite galaxies have higher galaxy-wide SFR-weighted metallicities than central galaxies, a phenomenon seen both in Illustris and in observations of real galaxies. We focus on a single mass scale, $M_*=(2-3)\times10^{10}\Msun$, and define several galaxy samples: centrals, satellites, two sub-samples of centrals and satellites that are closely matched in their formation histories, and a subset of satellites that form early and cannot be matched to a central in terms of formation history. We quantify the mean metallicity and SFR of these various samples as a function of galactocentric radius, cosmic time, and environment. Our investigations allow us to draw conclusions concerning what drives satellites to higher metallicities than centrals. Our main findings are summarized below.

\begin{itemize}
\item
Across a wide mass range, satellite galaxies in Illustris have higher metallicities than central galaxies, by $\approx0.15\dex$ at $M_*=10^9\Msun$ and $\approx0.1\dex$ at $M_*=10^{11}\Msun$ (\Fig{MZrelation}). Most of this difference is achieved by the satellite galaxies after their infall time onto their hosts, i.e., after becoming satellites (\Figs{samples_Z}{history_tinfall}).
\item
Satellite galaxies generally tend to form earlier than centrals. However, most ($\approx2/3$) satellite galaxies can be matched in terms of their formation history to a central galaxy (\Fig{samples}). The SFR profile, however, differs between these two matched populations, with satellite galaxies having more concentrated SFR profiles (\Fig{profileSFR}). This difference in SFR concentration is a main driver of inferred metallicity differences, as described below. As such, the scenario described here based on the Illustris simulation can be tested observationally by examining the existence of such differences, or lack thereof, between mass-matched centrals and satellites in the real Universe.
\item
The metallicity profiles (local metallicity as a function of radius) of formation-matched satellites and centrals are almost identical (\Fig{profileZlocal}). Based on a new definition of metallicity -- the `radially averaged' metallicity -- that is not affected by the SFR profile, the metallicities of formation-matched satellites and centrals are essentially the same (\Fig{Z_SFR_definitions}).
\item
All of the galaxy samples have mean negative metallicity gradients, i.e. they have higher metallicities at smaller radii (\Fig{profileZlocal}). It is the combination of this fact with the fact that the SFR profile is more concentrated in satellite galaxies that makes their overall SFR-weighted metallicities higher than those of centrals; it is not that the metallicity profiles themselves are different (\Fig{Z_SFR_definitions}). This weighting effect accounts for most ($\approx50-70\%$) of the difference in the SFR-weighted metallicities between satellites and centrals (\Fig{MZhistory}).
\item
The rest of the difference between the full satellite and central populations can be accounted for by their non-matched formation histories (\Figs{MZhistory}{Z_SFR_definitions}). First, a significant population of satellite galaxies forms early and has a low SFR at $z=0$ (\Fig{samples}). By virtue of the general metallicity-SFR anti-correlation (\Fig{Z_SFR}), this implies that satellites tend to have higher metallicities (\Fig{samples_Z}). However, there also exists an additional small effect wherein, for a given $z=0$ SFR, satellite galaxies that have early formation histories tend to have higher radially averaged (i.e., local) metallicities (\Figs{Z_SFR_definitions}{stellarage_SFR}). This may be explained by two effects caused by their older stellar populations: (i) higher metal mass release back to the galactic gas, and (ii) lower history-integrated wind mass-loading factors, implying less metal escape outside of the galaxy.
\item
Satellite galaxies in Illustris tend to have higher SFR-weighted metallicities in denser environments, as observed in reality (\Figs{Z_env}{Peng}). For satellites that are matched in formation history to centrals, this still holds for SFR-weighted metallicities (\Figs{Z_env_3kpc}{Z_env_30kpc}), but there is no relation for radially averaged metallicities with environment (\Fig{Z_env_radavg}). This suggests that what drives the relation with environment is the increased efficiency in denser environments of the SFR profile concentration/truncation that satellites undergo. A corollary is that the correlation with environment is weaker when the metallicity inside smaller galactocentric radii is concerned, a prediction that can be tested in the future in observational data (\Fig{Peng}).
\item
Some of the relation with environment that satellites present comes from the `non-matched' sample, and appears not only in the SFR-weighted metallicity but also in the radially averaged metallicity (\Fig{Z_env}). This relation appears because those satellites that are in denser environments have earlier formation histories (\Fig{stellarageSFR_environment}), and earlier formation histories imply higher metallicities for several reasons described above.
\item
We find that the circumgalactic gas around satellite galaxies tends to have higher metallicities than around centrals (\Fig{environment}). While the difference is (smaller yet) comparable to the absolute metallicity values of the star-forming gas inside the galaxies themselves, it is not the driver of the higher metallicities of the satellites with respect to the centrals. Future work will need to explore in more detail the metallicity of the accreting gas to understand this result.
\end{itemize}

\acknowledgements

We thank Erica Nelson and Jabran Zahid for useful discussions, and Christian Maier for assistance in incorporating his results for comparison with ours. We are particularly grateful to Nicolas Bouch{\'e} for insightful discussions and valuable comments on early drafts of this manuscript, and to the anonymous referee for a very thoughtful report. We are also indebted to Vicente Rodriguez-Gomez for providing galaxy merger histories, and to Mark Vogelsberger for making available measurements of environmental overdensities and mean stellar ages. The Illustris simulation was run on the CURIE supercomputer at CEA/France as part of PRACE project RA0844, and the SuperMUC computer at the Leibniz Computing Centre, Germany, as part of project pr85je. Post-processing computations used in this paper were run on the Odyssey cluster supported by the FAS Division of Science, Research Computing Group at Harvard University. SG acknowledges support provided by NASA through Hubble Fellowship grant HST-HF2-51341.001-A awarded by the STScI, which is operated by the Association of Universities for Research in Astronomy, Inc., for NASA, under contract NAS5-26555.

\label{lastpage}

\end{document}